\documentclass{article}
\usepackage{graphicx}
\usepackage[utf8]{inputenc}
\usepackage{lmodern}
\usepackage[T1]{fontenc}
\usepackage{lscape}
\usepackage{amsmath}	
\usepackage{amssymb}
\usepackage{cases}
\usepackage{graphicx}
\usepackage{caption}
\usepackage{comment}

\usepackage{makecell}
\usepackage{multirow}
\usepackage{amsfonts}
\usepackage{float} 
\usepackage{diagbox}
\usepackage{fullpage}
\usepackage{url}
\usepackage{rotating}
\usepackage{eurosym}
\usepackage{wrapfig}
\usepackage[final]{pdfpages} 
\usepackage{epstopdf} 
\usepackage[a4paper]{geometry}
\usepackage[nottoc]{tocbibind}
\usepackage{placeins}
\usepackage{float}
\usepackage{listings}
\usepackage{color, colortbl}
\usepackage{hyperref}
\usepackage{xcolor}
\usepackage{graphicx, caption, subcaption}

\usepackage{sidecap}
\hypersetup{
colorlinks,
citecolor=black,
filecolor=black,
linkcolor=black,
urlcolor=black
} 

\newcommand{\rtwo}{\textcolor{black}}
\newcommand{\rone}{\textcolor{black}}
\newcommand{\rot}{\textcolor{black}}

\newcommand\cute{CUTE }
\newcommand\pub{PUB }

\begin{document}

\title{Background Error Covariance Iterative Updating with Invariant Observation Measures for Data Assimilation
}

\author{Sibo Cheng$^{1,2}$, Jean-Philippe Argaud$^{1}$, Bertrand Iooss$^{1,3}$,  Didier Lucor$^{2}$, Angélique Pon{ç}ot$^{1}$ \\
        \small $^{1}$ EDF R\&D \\
        \small $^{2}$ LIMSI, CNRS, Université Paris-Saclay\\
        \small $^{3}$ Institut de Mathématiques de Toulouse, Université Paul Sabatier\\
}

%\date{\today }
\date{}

\maketitle

\begin{abstract}
In order to  leverage  the information embedded in the background state and observations, covariance matrices modelling is a pivotal point in data assimilation algorithms. These matrices are  often estimated from an ensemble of observations or forecast differences. Nevertheless, for many industrial applications the modelling still remains empirical based on some form of expertise and physical constraints enforcement in the absence of historical observations or predictions. We have developed two novel robust adaptive assimilation methods named  \cute (Covariance Updating iTerativE) and \pub (Partially Updating BLUE). {These two non-parametric methods are based on different optimization objectives, both} capable of  sequentially adapting background error covariance matrices in order to improve assimilation results under the assumption of a good knowledge of the observation error covariances. We have compared these two methods with the standard  approach using a misspecified background matrix in a shallow water twin experiments framework  with a linear observation operator. Numerical experiments have shown that the  proposed  methods bear  a real advantage both in terms of posterior error correlation identification and assimilation accuracy.

\end{abstract}

\section{Introduction}
Data assimilation methods are widely used in engineering applications with the objective of state-estimation or/and parameter registration/identification based on the \rtwo{weighted} combination of different sources of noisy information. \rtwo{Data assimilation often starts from some initial (i.e. {\em prior}) knowledge of the quantity of interest, and then produces a subsequent (i.e. {\em posterior}) estimator of it. Because of the noise alteration, it is most convenient if the method also provides some statistical information about the posterior estimator, at best in the form of its probability distribution.} These algorithms are very well known in geosciences and are used as  reference methods in the fields of numerical weather prediction (\cite{F.Parrish1992}), nuclear safety (\cite{Mei2017}), atmospheric chemistry (\cite{Singh2011}), \rot{hydrologic modelling \cite{Li2016}}, seismology, glaciology, agronomy, etc. Over decades, these approaches have been applied in the energy industry, for projects involving temperature field reconstruction (\cite{Argaud2014}) or forecasting in neutronic (\cite{Poncot2013}) and hydraulic (\cite{Goeury2017}). More recently, they have also made their way to other fields such as medicine, biomedical applications (\cite{lucor_esaim2018}) or wildfire front-tracking problems (\cite{rochoux_esaim2018}).

Data assimilation methods are based on prior estimation of the true state (also called background state) and one or several vectors of observations. There exist non-negligible errors in these two quantities. The essential idea is  therefore to find a compromise by fusing the information presented in these two quantities (\cite{carrassi2017}), while accounting for errors, in order to improve the quality of  field reconstruction and forecasting by learning from observations.  Due to lack of  knowledge, the background state is often provided by  some experts or approximated, e.g. from a numerical simulation.  A remarkable difficulty in the efficiency of these methods is that the prior error  covariance  matrices  are themselves imperfectly known, especially the one of background errors (often noted as matrix $\textbf{B}$). 

The modelling of $\textbf{B}$ as well as the observation error covariance matrix $\textbf{R}$, remains a very critical point in data assimilation problems because it determines how prior errors spread spatially  or \rone{temporally (e.g \cite{Senegas2001})} and this may substantially change  the assimilation results  (\cite{Hodges2010}). It also provides an important information of the relationship between observations and forecasts. As mentioned by \cite{Fisher2003}, there exist a wide variety of methods to estimate these matrices. Well known methods among others, are the one of  \cite{Hollingsworth1989b}, the \textit{NMC} \rone{(National Meteorological Center)} method (\cite{F.Parrish1992}) and ensemble methods (\cite{Clayton2012}) which are often combined with algebraic operations such as matrix factorization (\cite{Ishibashi2015}) or covariance localization (\cite{LIU2016}). For  many industrial applications, the paucity of historical observations as well as the large dimension and complexity (\rot{e.g.\cite{Sinsbeck2015}}) of the system, make the estimation of these covariance matrices unfeasible. A common practice in this case is to impose a standard form for the covariance matrices by empiricism. Certain types of matrices with homogeneous and isotropic characteristics such as diagonal matrices (\cite{Brian2005}) or relying on  generic  covariance kernels: e.g. Matérn kernel (\cite{Singh2011}), are often favoured. Other approaches are based on numerical techniques involving convolution operations (\cite{Gaspari1998}) or the resolution of diffusion equations (\cite{Weaver2013}). The latter methods are sometimes equivalent to the former ones, under simplifying assumptions, as explained in \cite{Mirouzethse}. 
An {\em a priori} choice of fixed covariance matrices with certain regularity properties thus  implicitly imposes extra assumptions to the problem, which may lead to supplementary uncertainties. Research efforts are continuously made toward improving the estimation of error covariances. 
Suffering from problems dimension and complexity, a variety of background matrix computation methods have been proposed in reduced spaces such as the spectral space (\cite{Courtier1998}) or the wavelet space (\cite{Chabot2017}). However, these contributions require prior assumptions about the matrix structure which could be difficult to justify in industrial applications. 

Recent works of \cite{Dreano2017}  have  also investigated model error covariance modelling (often noted as matrix $\textbf{Q}$), which can be seen as the main contributor of the background matrix in a dynamical system. Another pathway of research is to make assimilation more robust to this unavoidable lack of knowledge. This challenge applies
to both background and observation error covariances. In
this work, we focus on the former but both are important.

In this paper, we are interested in iterative algorithms that  can  be resilient to inconsistent prior background error covariance. More specifically, we look for algorithms that would  automatically adjust, thanks to an optimization process, the structure of the error covariance matrices. Our objective is to gain a better knowledge of error correlation which leads to a reduction of \textit{a posteriori} reconstruction errors with limited available data. The meteorology community has been a strong contributor to this topic, and several algorithms and their improved versions have been developed in \cite{Desroziers01}, \cite{Desroziers2005}, \cite{Chapnik2006b} etc. These methods have been applied world-widely in industrial problems ever since, e.g. \cite{Alistair2010}, \cite{Waller2016}. Among them, the Desroziers \& Ivanov tuning method  consists in finding a fixed point for the assimilated state by regulating the ratio between background and observation covariance matrices  magnitude without modifying their structures. This method, for which no statistical estimation of full matrices is required, could have a particular interest for industrial applications with limited prior data. However, it relies on a good  knowledge of the correlation of prior errors as shown in \cite{Chapnik2006b}. This last condition can be difficult to fulfil without enough historical statistics or in the case of a new application. \rtwo{Another iterative method, based on a diagnostic in the observation space (\cite{Desroziers2005}), aims at estimating the whole covariance matrices (see \cite{Janjic2018}). However, this method strongly relies on the statistics of either redundant observation data or historical innovation quantity, which are difficult to obtain in our industrial context. }
 
In this work, we develop two novel methods, consisting in repeating  several  times the assimilation procedure of the state-estimating problem with the same  set of  observations. A related idea of reusing several times the same observation data set has been carried out by \cite{Kalnay2010} in the \rone{"running in place"(RIP)} method  for the Ensemble Kalman Filtering in order to improve the system spin-up. We provide different approaches which are directly based on static Best Unbiased Linear Estimator (BLUE) and involve an updating of the background covariance matrix at the end of each iteration. We further take into account the covariance between the errors of  the updated background vectors and the ones of observations; this covariance appearing due to the iterative process itself. 
Based on this idea, we propose two iterative algorithms: \cute (Covariance Updating iTerativE) method and \pub (Partially Updating BLUE) method. These  methods can also be considered as some kind of preliminary  step, improving  a sequence of dynamical reconstruction with prediction, as they provide a more "consistent" covariance estimate, right after the first assimilation step. For numerical testing, a two-dimensional shallow water model with periodic boundary conditions is used to perform twin experiments for validation purposes. \rtwo{Both iterative methods are studied for a static reconstruction problem and  a dynamical data assimilation chain.}

The paper is organized as follows. Variational data assimilation is introduced briefly in section \ref{sec : Data assimilation and variational methods framework}, with a focus on covariance updating. We then propose two novel iterative methods in section \ref{sec:Iterative methods with advanced covariance updating},  together with a simple illustrative scalar test case. In section \ref{sec:numeric}, these methods are then compared on a two-dimensional fluid mechanics system in a twin experiments framework  for both state-independent and state-dependent prior errors.

\section{Data assimilation and variational methods framework}
\label{sec : Data assimilation and variational methods framework}
\subsection{Data assimilation concept} \label{subsec:Brief introduction to data assimilation}
The idea behind data assimilation system is to combine different sources of information in order to provide a more reliable estimation of the system state variables which can be \rone{a discretized physical field or a set of parameters (see \cite{Leisenring2011}}). We focus on the former where the state is presented by a vector of real entries which could for instance represent a discretized multidimensional physical field (e.g. speed, temperature) at some given coordinates. The true state is denoted by $\textbf{x}_t$. In general, the information is split into two parts: an initial state estimation $\textbf{x}_b$ (so called the background state)  and  an observation vector $\textbf{y}$, related to the state and representing  measurements. Both parts are noisy and the observations are often sparse especially for field reconstruction/prediction problems.  The observation operator $\mathcal{H}$ from the state space to the observable space is supposed to be known. Both background state and observations are uncertain quantities. Their tolerance, regarding theoretical (or `true') values, are quantified by $\epsilon_b$ and $\epsilon_y $, respectively:
\begin{align}
    \epsilon_b&= \rone{\textbf{x}_b-\textbf{x}_t}\\
    \epsilon_y&=\textbf{y}-\mathcal{H}(\textbf{x}_t). \notag
\end{align}

\rone{The transformation operator $\mathcal{H}$  and the true state $\textbf{x}_t$ are assumed to be deterministic quantities, except in the case of a dynamical data assimilation chain, which will not be discussed in great details in this paper. Following unbiased Gaussian distributions with covariance matrices $\textbf{B}$ and $\textbf{R}$, the background error $\epsilon_b$ and the observation error $\epsilon_y$ are supposed to be uncorrelated i.e}

\begin{align}
    &\epsilon_b \sim \mathcal{N}(0, \textbf{B})\\
    &\epsilon_y \sim \mathcal{N}(0, \textbf{R}) \notag\\
    COV(&\epsilon_b,\epsilon_y) = 0. \notag
\end{align}

\rot{Simply speaking, the inverse of these covariance matrices (i.e. $\textbf{B}^{-1}, \textbf{R}^{-1}$)} acts as some ``weights'' given to the different information sources.   In fact, these covariances not only describe the variation of estimation/instrument errors but also how they are correlated. \rone{These correlations may depend on the spatial distance, time scale or other physical quantities between two state variables or measure points.}

\rone{Things become slightly more complicated, due to the iterative approach proposed in this work to finely tune covariances. Indeed, some error correlation between {\em updated} state variables and observations may be induced by the iterative process. Therefore, it is crucial to account for this modified covariance, which will be discussed in full details later, in particular in section \ref{sec:Iterative methods with advanced covariance updating}. }

This approach is applied in a large variety of scientific domains, such as weather prediction, geophysical problems, signal processing, control theory etc. The mathematical handlings of data assimilation are mainly two-fold: Kalman filter-type methods based on estimation theory and  variational methods related to control theory.   Certain equivalences exist between these two families, especially when the transformation operator $\mathcal{H}$ is linear. \rot{Both approaches can be derived from Bayes' theorem (see \cite{carrassi2017})} where the  state estimation $\textbf{x}_a$ provided by the data assimilation procedure may be apprehended as a compromise between the information of background estimation and the ones of observations. \rot{In practice,  dealing with nonlinear problems of large dimension via Bayesian approaches remains a computationally expensive task. } \rot{In this paper}, we focus on the framework of \rone{linearized} variational methods. However, the analysis and algorithms developed later in this paper can be directly applied  at each updating of Kalman filter-type methods.

\subsection{Variational formulation} \label{sec:Introduction to 3D-Var}

In order to better focus on the study of background covariance matrix computation, we suppose in this paper that $\mathcal{H}$ is linear and perfectly known, represented in matrix form by $\textbf{H}$ from now on. For this reason, we refer to instrument errors when computing the observation matrix $\textbf{R}$. In the case of field reconstruction, the observation $\textbf{y}$ is only supposed to provide a partial information on the true state.   

As mentioned in section \ref{subsec:Brief introduction to data assimilation}, the key idea in variational methods is to find a balance between the background and the observations (\cite{Bou1999}) according to the weights represented by \rone{the inverse} of $\textbf{B}$ and $\textbf{R}$. This leads to the loss function:  
\begin{align}
    J_{\text{3D-VAR}}(\textbf{x})&= \frac{1}{2}(\textbf{x}-\textbf{x}_b)^T \textbf{B}^{-1}(\textbf{x}-\textbf{x}_b) + \frac{1}{2}(\textbf{y}-\mathcal{H}(\textbf{x}))^T \textbf{R}^{-1} (\textbf{y}-\mathcal{H}(\textbf{x})) \label{eq_3dvar}\\
   &=\frac{1}{2}||\textbf{x}-\textbf{x}_b||^2_{\textbf{B}^{-1}}+\frac{1}{2}||\textbf{y}-\mathcal{H}(\textbf{x})||^2_{\textbf{R}^{-1}} \label{eq:op_3dvar}.
\end{align}
The optimisation problem defined by the objective function of Eq.  (\ref{eq:op_3dvar})  is  called three-dimensional variational method (\textit{3D-VAR}), which can also be considered as the general equation of variational methods \rtwo{without considering the transition model error (i.e. except weak-constraint data assimilation) (see.\cite{carrassi2017})}.

\subsection{Best Linear Unbiased Estimator (BLUE)} \label{sec:BLUE equivalence}

\rot{Given some observed datasets (in our case both $\textbf{x}_b$ and $\textbf{y}$) and the associated error variance, the Best Linear Unbiased Estimator (BLUE) combines both source of information to produce an unbiased linear estimator with minimum posterior variance (see \cite{Asch2016}). }
\rot{When $\mathcal{H} = \textbf{H}$ is linear,} the {\em optimal} solution provided by the variational formulation (Eq.\ref{eq:op_3dvar}) is identical to the one obtained by a BLUE under the assumption of independence between $\textbf{x}_b$ and $\textbf{y}$ in terms of prior estimation errors. This approach is unbiased and minimises optimally the error variance, {\em assuming the error covariance matrices are perfectly known}. It also coincides with the maximum likelihood estimator when prior errors of both background state and observations are normally distributed. In this case, the analysed state $\textbf{x}_a$ can be updated explicitly as: 
\begin{equation}
	\textbf{x}_a=\textbf{x}_b+\textbf{K}(\textbf{y}-\textbf{H} \textbf{x}_b) \label{eq:BLUE_1}
\end{equation}
where the Kalman gain matrix $\textbf{K}$ is defined as:
\begin{equation}
	\textbf{K}=\textbf{B} \textbf{H}^T (\textbf{H} \textbf{B} \textbf{H}^T+\textbf{R})^{-1}. 
	\label{eq:Kgain_BLUE}
\end{equation}

Under the assumption of linearity, the covariance of analysis error  $\epsilon_a=(\textbf{x}_a-\textbf{x}_t)$ takes an exact explicit form:
\begin{align}
\textbf{A}&= COV(\textbf{x}_a-\textbf{x}_t) \notag\\
&=(\textbf{I}-\textbf{K}\textbf{H}) \textbf{B} (\textbf{I}- \textbf{K} \textbf{H})^T + \textbf{K} \textbf{R} \textbf{K}^T \notag \\
&=(\textbf{I}-\textbf{K}\textbf{H})\textbf{B}. \label{eq:A_a}
\end{align}

\rone{Under the assumption that both background and observation errors follow centered Gaussian distributions, it is easy to justify  that:}
\begin{align}
    \epsilon_a =  \textbf{x}_a - \textbf{x}_t \sim \mathcal{N} (0, \textbf{A}). \label{eq:eps_a}
\end{align}

\rone{Eq.\ref{eq:eps_a} holds when $\textbf{H}$ is linear. We recall that the uncorrelatedness between $\epsilon_b$ and $\epsilon_y$ is a crucial assumption for Eq.\ref{eq:BLUE_1}-\ref{eq:eps_a}. Furthermore, the assumption of Gaussianity on prior errors ensures the complete knowledge of posterior errors distribution as a Gaussian vector can be fully represented by its expectation and covariance. Nonetheless, the estimation of posterior covariance in Eq.\ref{eq:A_a}, as well as in the iterative tuning methods proposed in this work remains valid as long as the prior information (both $\textbf{x}_b$ and $\textbf{y}$) is unbiased, regardless of the nature of prior distributions. A more general form of BLUE is presented latter in section \ref{sec:PUB}. Non-Gaussianity in data assimilation problems, for example due to the nonlinearity of $\mathcal{H}$, has been discussed (e.g \cite{Sorensen2004}). }\\
However, we emphasize that when prior covariances are not well known, the  estimation provided by Eq.\ref{eq:A_a} could be very different from the exact\footnote{Here, by the term ``exact'', we refer to the covariance truly corresponding to the remaining errors present in the analysed state, no matter the level of optimality of the chosen assimilation scheme.} output error covariance (later noted as $\textbf{A}_\textsc{E}$). It is therefore of highest importance to differentiate between well or loosely known prior covariance matrices. 
This aspect will be investigated further in section \ref{sec:Iterative methods with advanced covariance updating}.

\subsection{Misspecification of  $\textbf{B}$ matrix}
\label{subsec:mis B}
From here, we follow the notations given in \cite {Eyre2013}, where $\textbf{B}_\text{E}$ designates the unknown exact background error covariance while $\textbf{B}_{\text{A}}$ stands for the {\em assumed} (or guessed) matrix which can be considered as a parametric quantity within data assimilation algorithms. In this section, we focus on the impact on the output error covariance and its estimation given by the misspecification of matrix $\textbf{B}_\text{E}$ (mismatch between $\textbf{B}_\text{E}$ and $\textbf{B}_\textsc{A}$).
Following the current notation, the standard estimation of output error covariance $\textbf{A}_{\textsc{A}}$ (here we have kept subscript A to indicate that this form is obtained from $\textbf{B}_{\text{A}}$) provided by the plain-vanilla \textit{3D-VAR} method in Eq. (\ref{eq:A_a}) becomes:
\begin{align}
    \textbf{A}_{\textsc{A}}=(\textbf{I}-\textbf{K}(\textbf{B}_\textsc{A})\textbf{H})\textbf{B}_\textsc{A}, \label{eq:assumed_A}
\end{align}
which is different from the exact output error covariance when the unknown $\textbf{B}_\text{E}$ has merely been approximated by $\textbf{B}_{\textsc{A}}$. The gain matrix $\textbf{K}$ remains a function of the assumed background covariance matrix $\textbf{B}_{\textsc{A}}$ and the one of the observation, $\textbf{R}$. The latter is supposed to be perfectly known. In fact, the exact output error covariance $\textbf{A}_\text{E}$ depends on the prior error and the parameters of the algorithm. Therefore, as described by \cite{Eyre2013}, $\textbf{A}_\text{E}$ is in function of $\textbf{B}_\text{E}$ and $\textbf{B}_\textsc{A}$:   
\begin{align}
    \textbf{A}_\text{E}=(\textbf{I}-\textbf{K}(\textbf{B}_{\textsc{A}})\textbf{H}) \textbf{B}_\text{E} (\textbf{I}-\textbf{K}(\textbf{B}_{\textsc{A}})\textbf{H})^T + \textbf{K}(\textbf{B}_{\textsc{A}}) \textbf{R} \textbf{K}(\textbf{B}_{\textsc{A}})^T. \label{eq:true_A}
\end{align}

As we have mentioned before, the final analysis of the assimilation procedure is very much dependent on the specification of the weights given to background and observations, through the error covariances. \rtwo{In fact, when the background matrix is perfectly specified, i.e. $\textbf{B}_\textrm{A} = \textbf{B}_\textrm{E}$, the obtained Kalman gain matrix $\textbf{K} (\textbf{B}_\textrm{E})$  is a so called {\em optimal} gain matrix, which ensures that the trace of $\textbf{A}_\textrm{E}$ is minimal. }
Because the covariance of these errors are not well known, it is natural to turn to methods producing {\em a posteriori} diagnoses of the misspecification of the {\em a priori} errors, in order to (sequentially) adapt them. For instance, the Desroziers tuning algorithms (\cite{Desroziers01}) allow the adjustment of the multiplicative ratio (i.e. the total variance) between matrices $\textbf{B}_{\textsc{A}}$ and $\textbf{R}$, in order to improve the quality of the analysis.

Our goal is somewhat different from the Desroziers tuning algorithm, as we wish to gain a better knowledge about the error correlation pattern/structure of the output analysis. Indeed, the knowledge of error correlation is crucial for posterior analysis and provides a finer information than the error variance alone. We remind in general how a covariance matrix $\textbf{Cov}$ is related to its correlation matrix $\textbf{Cor}$:
\begin{align}
    \textbf{Cov}=\textbf{D}^{\frac{1}{2}} \textbf{Cor} \textbf{D}^{\frac{1}{2}} \label{eq:correlation_covariance_2}
\end{align}
where $\textbf{D}$ is a diagonal matrix with identical diagonal elements of $ \textbf{Cov}$ and thus $\textbf{D}^{\frac{1}{2}}$ represents the standard deviations.  

\subsection{Data assimilation for dynamical systems}
\label{DA dynamical}
\rone{Data assimilation algorithms could be applied to dynamical systems thanks to a sequential application of variational methods using a transition operator (from discretized  time $t^k$ to $t^{k+1}$) $\mathcal{M}_{t^k \rightarrow t^{k+1}}$, where}
\begin{align}
    \textbf{x}_{t^{k+1}} = \mathcal{M}_{t^k \rightarrow t^{k+1}} (\textbf{x}_{t^k }).
\end{align}
\rone{The forecasting in data assimilation thus relies on the knowledge of transition operator  $\mathcal{M}_{t^k \rightarrow t^{k+1}}$ and the corrected state at the current time $ \textbf{x}_{a, t^k }$. The state correction could be carried out at each time step $t=t^k$ with current observation $\textbf{y}_{t^k}$. Typically, the background state is often provided by the forecasting from the previous step, i.e.}
\begin{align}
    \textbf{x}_{b,t^k} = \mathcal{M}_{t^{k-1} \rightarrow t^{k}} (\textbf{x}_{a,t^{k-1} }).
\end{align}
\rone{It is known that as long as the transformation operator $\mathcal{H}$ and the transition operator $\mathcal{M}$ are linear, the analysis based on the variational method (\textit{4D-VAR}) and the Kalman filter leads to the same forecasting result (\cite{Fisher2005}). As in both cases the approximation of $\mathcal{M}$ may bring extra noises which may probably lead to a nonlinear error propagation, we think an error covariance diagnostic/correction at different time step could be helpful. Not relying on the dynamic of the system, the iterative tuning methods proposed in this paper could be applied at any step in a data assimilation chain.} 

\section{Iterative variational methods with advanced covariance updating} \label{sec:Iterative methods with advanced covariance updating}
For interpolation of complex industrial applications, the model error, \rone{due to the approximation of the transition model $\mathcal{M}$, } is often integrated as a part of the background error. This modelling choice usually leads to a less precise knowledge about the background covariance matrix $\textbf{B}$ relative to the observation covariance matrix $\textbf{R}$. Therefore, we consider that the background errors are dominant over observation errors with a noise-free transformation operator $\mathcal{H}$, but the exact ratio between them is difficult to estimate.  It was pointed out in \cite {Eyre2013} that an {\em overestimation} of covariance $\textbf{B}$ will introduce a significant risk of mis-calculating the output error covariances. \rone{As a consequence, the main idea of our iterative methods is to iterate the data assimilation procedure for a better posterior state estimation and error covariance specification, avoiding overestimation of $\textbf{B}$. Therefore, the adjustment of the state variables and its covariance associated will take place progressively. } 

\subsection{Naive approach}
 In practice, the data assimilation procedure can be reapplied several times making use of the same observations, in order to balance the weight between background states and observations. This naive approach may be summarised as:
\begin{eqnarray}
    \textbf{x}_{b,n+1} & \leftarrow & \textbf{x}_{a,n}=\textbf{x}_{b,n}+\textbf{K}_n(\textbf{y}-\textbf{H} \textbf{x}_{b,n}) \label{eq:naive1} \\ 
    \textbf{B}_{\textsc{A},n+1}& \leftarrow & \textbf{A}_{\textsc{A},n}= (\textbf{I}-\textbf{K}_{n}\textbf{H}) \textbf{B}_{\textsc{A},n}, \label{eq:naive2}
\end{eqnarray}
 where $n$ refers to the iteration number, and:
\begin{align}
    \textbf{K}_n=\textbf{B}_{\textsc{A},n}\, \textbf{H}^T (\textbf{H} \textbf{B}_{\textsc{A},n} \textbf{H}^T+\textbf{R})^{-1},
    \label{eq:Kn}
\end{align}
is the iterated Kalman gain matrix.

\subsection{Mis-calculation of updated covariances}
\label{sec:Mis-naive}
\rone{The updating of error covariances is incorrect because of the state-observation error correlation emerging due to the iterative process}. \rone{In fact, the evolution of the exact analysed/background error covariance $\textbf{A}_{E,n}$/$\textbf{B}_{E,n}$ can be expressed as a function of $\textbf{B}_{\textsc{A},n}$ and $\textbf{K}_n$}:

\begin{align}
     \textbf{B}_{\textsc{E},n+1}&=\textbf{A}_{\textsc{E},n}=(\textbf{I}-\textbf{K}_{n}\textbf{H}) \textbf{B}_{\text{E},n} (\textbf{I}-\textbf{K}_{n} \textbf{H})^T +(\textbf{I}-\textbf{K}_{n} \textbf{H})\textbf{Cov}(\epsilon_{b,n},\epsilon_y) \textbf{K}_{n}^T \notag\\ 
&+\textbf{K}_{n} \textbf{Cov}(\epsilon_y, \epsilon_{b,n},) (\textbf{I}-\textbf{K}_{n}\textbf{H})^T+\textbf{K}_{n} \textbf{R} \textbf{K}_{n}^T, \label{eq:itrative_3dvar}
\end{align}

\noindent where $\textbf{Cov}(\epsilon_{b,n},\epsilon_y)=\textbf{Cov}(\epsilon_y, \epsilon_{b,n})^T$ represents the error covariance of $\textbf{x}_{b,n}$ and $y$. \rone{Indeed, the state errors are no longer uncorrelated to the observation ones after the first iteration, i.e.
\begin{align*}
    \textbf{Cov}(\epsilon_{b,n},\epsilon_y) \neq 0 \quad \textrm{for} \quad \forall n \geq 1.
\end{align*}
}  As a result, the exact analysis error covariance $\textbf{A}_{\textsc{E},n}$ tends to be under-estimated by $\textbf{A}_{\textsc{A},n}$ in Eq.\ref{eq:naive2} throughout the iterations.

This is an important drawback that we next attempt to illustrate in a straightforward scalar case, where we assume:
\begin{align}
    B_\textsc{A}, R \in \mathbb{R}^+\setminus \{0\}, H \in \mathbb{R}\setminus \{0\}.
\end{align}
Here, we keep the covariance matrix denomination for notation coherence but they only reflect scalar variances.\\
In this case,
\begin{align}
    B_{\textsc{A},n+1}& \leftarrow \Big(1-\frac{B_{\textsc{A},n} H^2}{B_{\textsc{A},n} H^2+R} \Big) B_{\textsc{A},n}=\frac{B_{\textsc{A},n} R}{B_{\textsc{A},n} H^2+R}.  \label{eq:update_naive1}
\end{align}
In fact, one may see that the assumed error covariance (scalar variance in this case) $\textbf{B}_{\textsc{A},n\rightarrow \infty}$ provided by the naive iterations converges to zero, therefore falsely suggesting a reasonable estimator. 
This convergence can be easily proved by studying the fixed-point and monotonicity of the function $f$ in $\mathbb{R}^+$ defined as:
\begin{align}
    f(x)=\frac{x R}{x H^2 + R}, \quad \textrm{where} \quad R \in  \mathbb{R}^+\setminus \{0\}, H \in \mathbb{R}\setminus \{0\}.
\end{align}

\noindent Zero is obviously the only fixed-point of $f$. On the other hand, 
\begin{align}
    \forall x \in \mathbb{R}^+\setminus \{0\},  \quad f(x)<x,
\end{align}

and $f^{(n)}(x)$ (\rone{$f^{(n)}(x) = f(f^{(n-1)}(x))$}) is a decreasing sequence with a lower bound zero, thus it is convergent. Because zero is the only fixed-point of $f$, we can conclude that $B_{\textsc{A},n\rightarrow \infty}\rightarrow 0$ for any initial value $B_{\textsc{A}} \in \mathbb{R}\setminus \{0\}$.
This theoretical result is numerically confirmed in Fig \ref{fig:1d_analyse}, (solid green line). The distribution of the exact covariance, consistent with the updating loop (Eq. (\ref{eq:itrative_3dvar})), is depicted by the dashed green line and remains positive and non-zero.\\

Based on this idea, we propose two different algorithms named \cute and \pub, aiming at a better control of the output error correlation, and consequently a reduction of assimilation error.
From now on, for the simplicity of analysis, we make further hypothesis about the error covariance matrix $\textbf{R}$ of observations to be well known.

\subsection{\cute (Covariance Updating iTerativE) method} 

 \rone{As pointed out in section \ref{sec:Mis-naive}, the state-observation covariance $COV(\epsilon_{b,n}, \epsilon_y)$ must be taken care of in the covariance updating.}\\
\subsubsection*{Algorithm}
As we have mentioned in the previous sections, when $\mathcal{H}=\textbf{H}$ is a linear operator, the reconstructed state $\textbf{x}_a$ can be expressed as a linear combination of $\textbf{x}_b$ and $\textbf{y}$. Therefore, the covariance of updated background state and observations can be estimated sequentially as:

\begin{align}
\textbf{Cov}(\epsilon_{b,n},\epsilon_y)& = \textbf{Cov}(\epsilon_y, \epsilon_{b,n})^T = \textbf{Cov}\Big( \big[ (\textbf{I}-\textbf{K}_{n-1}\textbf{H})\epsilon_{b,n-1}+\textbf{K}_{n-1}\epsilon_y \big], \epsilon_y \Big)\\
& =(\textbf{I}-\textbf{K}_{n-1}\textbf{H}) \textbf{Cov}(\epsilon_{b,n-1}, \epsilon_y)+\textbf{K}_{n-1}\textbf{R}, \label{eq: COV_\cute}
\end{align}

\noindent with

\begin{equation}
    \textbf{Cov}(\epsilon_{b,0},\epsilon_y) = 0_{\textrm{dim}(x_b) \times \textrm{dim}(y)}.
\end{equation}
In practice, especially in the case of a poor quality of matrix specification \textit{a priori}, we have found that it is helpful to \rtwo{control} the trace of matrices $\textbf{B}_{\textsc{A},n}$ at each iteration in order to balance the weight of background state and observations. \rtwo{Indeed, if no care is taken of that, the norm of the updated covariance $\textbf{B}_{\textrm{A},n}$ may reduce too quickly after the first iteration, thereby causing a neglect of the observation data during the next iterations. Therefore a scaling is introduced through a coefficient $\alpha \in (0,1)$ related to the confidence level of prior matrix estimation, in order to control the trace of the updating matrix $Tr(\textbf{B}_{\textrm{A},n+1})$. 
The latter, representing the posterior covariance estimation, is introduced in Eq.\ref{eq: BA \cute}}.

The complete update of the state and the background covariance matrix is therefore written as:
\begin{align}
\textbf{x}_{b,n+1}&  \leftarrow \textbf{x}_{a,n},\\
\textbf{A}_{\textsc{A},n} & =(\textbf{I}-\textbf{K}_{n}\textbf{H})\textbf{B}_{\textsc{A},n} +(\textbf{I}-\textbf{K}_{n}\textbf{H})\textbf{Cov}(\epsilon_{b,n},\epsilon_y) \textbf{K}_{n}^T\\ \label{eq: BA \cute}
&+\textbf{K}_{n} \textbf{Cov}(\epsilon_y,\epsilon_{b,n}) (\textbf{I}-\textbf{K}_{n}\textbf{H})^T, \notag \\
\rtwo{\textbf{B}_{\textsc{A},n+1}}& \rtwo{\leftarrow  \frac{(1-\alpha) Tr(\textbf{B}_{\textsc{A},n}) + \alpha Tr(\textbf{A}_{\textsc{A},n})}{Tr(\textbf{A}_{\textsc{A},n})} \textbf{A}_{\textsc{A},n}}
\end{align}
where $\textbf{K}_{n}$ is expressed in Eq. (\ref{eq:Kn}). \rtwo{The more confident we are in the initial guess $\textbf{B}_{\textrm{A},0}$, the higher level of $\alpha$ should be set. In the extreme case where the initial background matrix is set arbitrarily (which is not rare in industrial applications), setting $\alpha = 0$ is suggested which means the trace of $\textbf{B}_{\textrm{A},n}$ will be kept constant in the iterative process. } 

\subsubsection*{Analysis}
It should be mentioned that, despite our effort on taking the background-observation covariance into account, the evolution of the error covariance can not be perfectly known due to the misspecification of $\textbf{B}$ at the first iteration. The evolution of exact analysis/background error covariance $\textbf{B}_{\textsc{E},n+1}$, which depend on the set up of $\textbf{B}_{\textrm{A},n}$, can be expressed as:
\begin{align}
       \textbf{B}_{\textsc{E},n+1} \leftarrow  \textbf{A}_{n} &=(\textbf{I}-\textbf{K}_{n}\textbf{H}) \textbf{B}_{\textsc{E},n} (\textbf{I}-\textbf{K}_{n}\textbf{H})^T +(\textbf{I}-\textbf{K}_{n}\textbf{H})\textbf{Cov}_\textsc{E}(\epsilon_{b,n},\epsilon_y) \textbf{K}_{n}^T\\
&+\textbf{K}_{n} \textbf{Cov}_\textsc{E}(\epsilon_y,\epsilon_{b,n}) (\textbf{I}-\textbf{K}_{n} \textbf{H})^T+\textbf{K}_{n} \textbf{R} \textbf{K}_{n}^T. \notag
\end{align}

We remind that the term $ \textbf{Cov}_\textsc{E},(\epsilon_{b,n},\epsilon_y)$, which represents the exact background-observation covariances, is calculated as done in Eq. (\ref{eq: COV_\cute}) but using the exact updated covariance matrix $\textbf{B}_{\textsc{E},n}$ in the expression of $\textbf{K}$.

Estimating the covariance introduced between $\textbf{x}_{b,n}$ and $\textbf{y}$, at each iteration, allows for a more "consistent" update, in the sense that if the estimation of $\textbf{B}$ and $\textbf{R}$ becomes asymptotically accurate, the iterative process will not add extra errors to the posterior covariance estimate. However, since the covariance between $\textbf{x}_b$ and $\textbf{y}$ emerges, the optimality of a \textit{3D-VAR}  formula in a loop ( Eq. (\ref{eq:op_3dvar})) may be questioned. Therefore, under the assumption of linearity, we propose another formulation that relies directly on the BLUE estimator in an extended space.

\subsection{\pub  (Partially Updating Blue) method} \label{sec:PUB}
 \rone{With the \cute formulation, we have taken $COV(\epsilon_{b,n}, \epsilon_y)$ into account in the covariance updating but they are not considered in the optimization loss function (Eq. (\ref{eq:op_3dvar})).}  \rone{To overcome this shortage}, our idea is to merge the background and observations in a broader space of larger dimension (as shown in \cite{Talagrand99}) with a partial updating dealing only concerning the part of the background state and its associated covariance.  \rone{By merging the state and the observation space, the cross-covariances $COV(\epsilon_b, \epsilon_y)$ could be taken into account in the iterative applications of minimisation problems using BLUE-type formulation.}\\

\subsubsection*{Algorithm}
In general, the BLUE estimator consists of constructing an unbiased estimate with minimum of variance from a true state $\theta$, an observation $z$, a transformation operator $\tilde{\textbf{H}}$ from the state to the observation space and the observation error covariance matrix $C$, under the assumption that:
\begin{equation}
\textbf{z}=\tilde{\textbf{H}} \theta + \textbf{w},
\end{equation}
where $\textbf{w}$ is a white noise.\\

The minimization of state minus observation under the norm defined by the error covariance $C$ is:
\begin{align}
   J(\textbf{x})=\frac{1}{2}||\textbf{z}-\tilde{\textbf{H}} \theta||_{C^{-1}}, \label{eq:global_blue}
\end{align}
and yields the BLUE $\hat{\theta}$ and its output covariance estimation $\textbf{C}_{\hat{\theta}}$:
 
\begin{align}
\hat{\theta}&=(\tilde{\textbf{H}}^T \textbf{C}^{-1} \tilde{\textbf{H}})^{-1} \tilde{\textbf{H}}^T \textbf{C}^{-1} \textbf{z}, \nonumber \\
\textbf{C}_{\hat{\theta}}&=(\tilde{\textbf{H}}^T \textbf{C}^{-1} \tilde{\textbf{H}})^{-1}. \label{eq: Blue_global}
\end{align}

\rtwo{Here we refer to the general form of the BLUE without any extra assumptions, for example, the uncorrelation between $\epsilon_b$ and $\epsilon_y$ as in section \ref{sec:BLUE equivalence}.  Furthermore, under the assumption of linearity of $\textbf{H}$, Eq.\ref{eq: Blue_global} is equivalent to the Maximum Likelihood Estimator.} In order to be well adapted to the general framework of the BLUE estimator, we redefine the system (\ref{eq:BLUE_1}-\ref{eq:A_a}) by simply combining the observation  and the background spaces:
\begin{align}
\theta \equiv \textbf{x}_t, \quad
\textbf{z} \equiv \begin{pmatrix}
        \textbf{x}_{b}  \\
		\textbf{y}  
    \end{pmatrix}, \quad
 \tilde{\textbf{H}} \equiv \begin{pmatrix}
        \textbf{I} \\
		\textbf{H}
    \end{pmatrix}, \quad  
    \textbf{w} \equiv \begin{pmatrix}
        \epsilon_{b}  \\
		\epsilon_y  
    \end{pmatrix}, \quad  
\textbf{C} \equiv \begin{pmatrix}
        \textbf{B} & 0  \\
		0 & \textbf{R}  
    \end{pmatrix}. \label{eq:PUB_matrix}
\end{align}

Similarly to the previous algorithms, we suppose that the matrix $\textbf{B}$ is misspecified, which yields also a misspecification of the matrix $\textbf{C}$ in Eq. (\ref{eq:PUB_matrix}). The assumed covariance matrix in the extended space is denoted as $\textbf{C}_\textsc{A}$ ($\textbf{C}_\textsc{A,0}$ latter in the iterative method) with no initial covariance between the background state and the observations, which can be written as:
\begin{align}
 \textbf{C}_\textsc{A} \equiv \begin{pmatrix}
        \textbf{B}_\textsc{A} & 0  \\
		0 & \textbf{R}
    \end{pmatrix}. \label{eq:PUB_matrix}   
\end{align}
As for the \cute method, we aim to adjust the structure of error covariance matrix $\textbf{C}_\textsc{A}$  by taking into account the covariance between the updated background  and the observation, which yields the updating loop of \pub method:
\begin{align}
\textbf{x}_{b,n+1}  \leftarrow \textbf{x}_{a,n}&=(\tilde{\textbf{H}}^T \textbf{C}_{\textsc{A},n}^{-1} \tilde{\textbf{H}})^{-1} \tilde{\textbf{H}}^T \textbf{C}_{\textsc{A},n}^{-1} \textbf{z}_n, \\
    \textbf{z}_{n+1}&=\begin{pmatrix}
        \textbf{x}_{b,n+1}  \\
		\textbf{y}  
    \end{pmatrix}\\
 \textbf{A}_{\textsc{A},n} &=(\tilde{\textbf{H}}^T \textbf{C}_{\textsc{A},n}^{-1} \tilde{\textbf{H}})^{-1}\\
 \rtwo{\textbf{B}_{\textsc{A},n+1}}& \rtwo{ \leftarrow  \frac{(1-\alpha) Tr(\textbf{B}_{\textsc{A},n}) + \alpha Tr(\textbf{A}_{\textsc{A},n})}{Tr(\textbf{A}_{\textsc{A},n})} \textbf{A}_{\textsc{A},n}}\\
 \textbf{Cov}(\epsilon_{b,n+1},\epsilon_y)&=(\tilde{\textbf{H}}^T \textbf{C}_{\textsc{A},n}^{-1} \tilde{\textbf{H}})^{-1}  \tilde{\textbf{H}}^T \textbf{C}_{\textsc{A},n}^{-1} \begin{pmatrix}
        \textbf{Cov}(\epsilon_{b,n},\epsilon_y)  \\
		R  
    \end{pmatrix}\\
    \textbf{C}_{\textsc{A},n+1}&=\begin{pmatrix}
        \textbf{B}_{\textsc{A},n+1} & \textbf{Cov}(\epsilon_{b,n+1},\epsilon_y)  \\ \label{eq:BA \pub}
	\textbf{Cov}(\epsilon_{b,n+1},\epsilon_y)^T & \textbf{R}  
    \end{pmatrix}
\end{align}

\noindent where $ \textbf{C}_{\textsc{A},n}$ is the {\em assumed} error covariance matrix in the combined space of background state and observations. \rtwo{Similar to \cute, the coefficient $\alpha$ is introduced to balance the ratio between {\em assumed} covariances in $ \textbf{C}_{\textsc{A},n}$}

\subsubsection*{Analysis}
Let $ \textbf{C}_{\textsc{E},n}$ denotes the exact iterated error covariance with $\textbf{Cov}_\textsc{E}(\epsilon_{b,n},\epsilon_y)$ representing the exact covariance between $\textbf{x}_{b,n}$ and $\textbf{y}$ in this extended space, i.e.
\begin{align}
        \textbf{C}_{\textsc{E},n}&=\begin{pmatrix}
        \textbf{B}_{\textsc{E},n} &\textbf{Cov}_\textsc{E}(\epsilon_{b,n},\epsilon_y)  \\ 
	\textbf{Cov}_\textsc{E}(\epsilon_{b,n},\epsilon_y)^T & \textbf{R} 
    \end{pmatrix},
\end{align}

\noindent where, as for the \cute method, we can also express the exact background  error covariance evolution as:

\begin{align}
    \textbf{B}_{\textsc{E},n+1} &=(\tilde{\textbf{H}}^T \textbf{C}_{\textsc{A},n}^{-1} \tilde{\textbf{H}})^{-1} \tilde{\textbf{H}}^T \textbf{C}_{\textsc{A},n}^{-1} \hspace{1mm} \textbf{C}_{\textsc{E},n} \hspace{1mm} \Big((\tilde{\textbf{H}}^T \textbf{C}_{\textsc{A},n}^{-1} \tilde{\textbf{H}})^{-1} \tilde{\textbf{H}}^T \textbf{C}_{\textsc{A},n}^{-1}\Big)^T  \label{eq:update cov_\pub}.
\end{align}
We note that this method does not only take into account the updated variances but also modify the optimisation formula (\ref{eq:global_blue}) in the extended space. This effect could make the \pub method more robust and less sensitive to prior assumptions, which will be shown later in section \ref{sec:numeric}.  However, the implementation of the algorithm, especially the matrix conditioning could be a bit more sophisticated due to the vector space of a larger dimension. 

\subsection{Comparison of these methods using an illustrative simple scalar case}

As we explained earlier, the objective of the proposed iterative methods is to obtain a better knowledge of the covariance (amplitude and/or correlation, depending on the application and prior knowledge) of output errors which can be crucial for future predictions in a data assimilation chain.

\begin{figure}[!ht]
  \centering
     \includegraphics[trim=2cm 8cm 4cm 9.5cm,clip=true,width=3.3in]{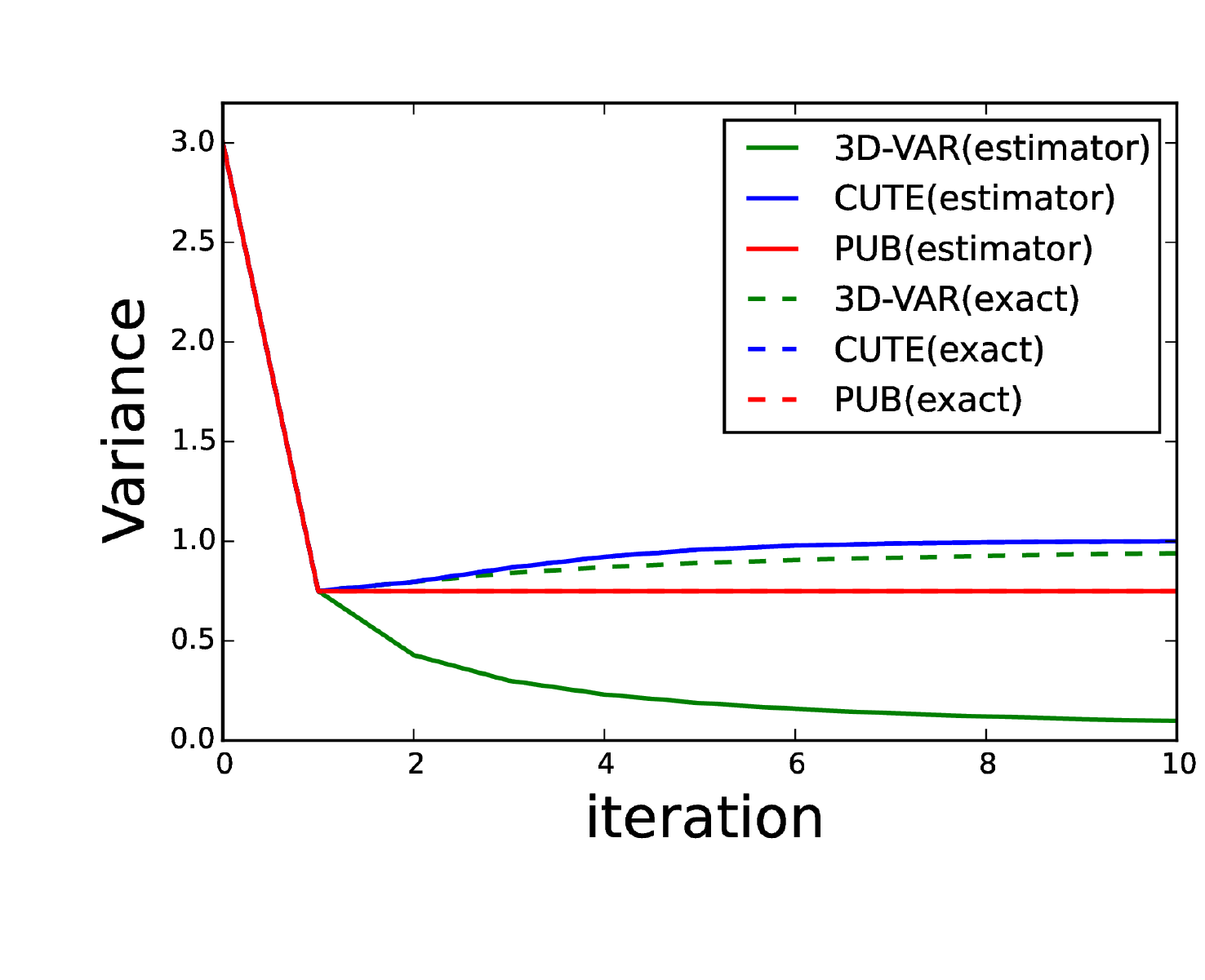}
    \includegraphics[trim=2cm 8cm 4cm 9.5cm,clip=true, width=3.3in]{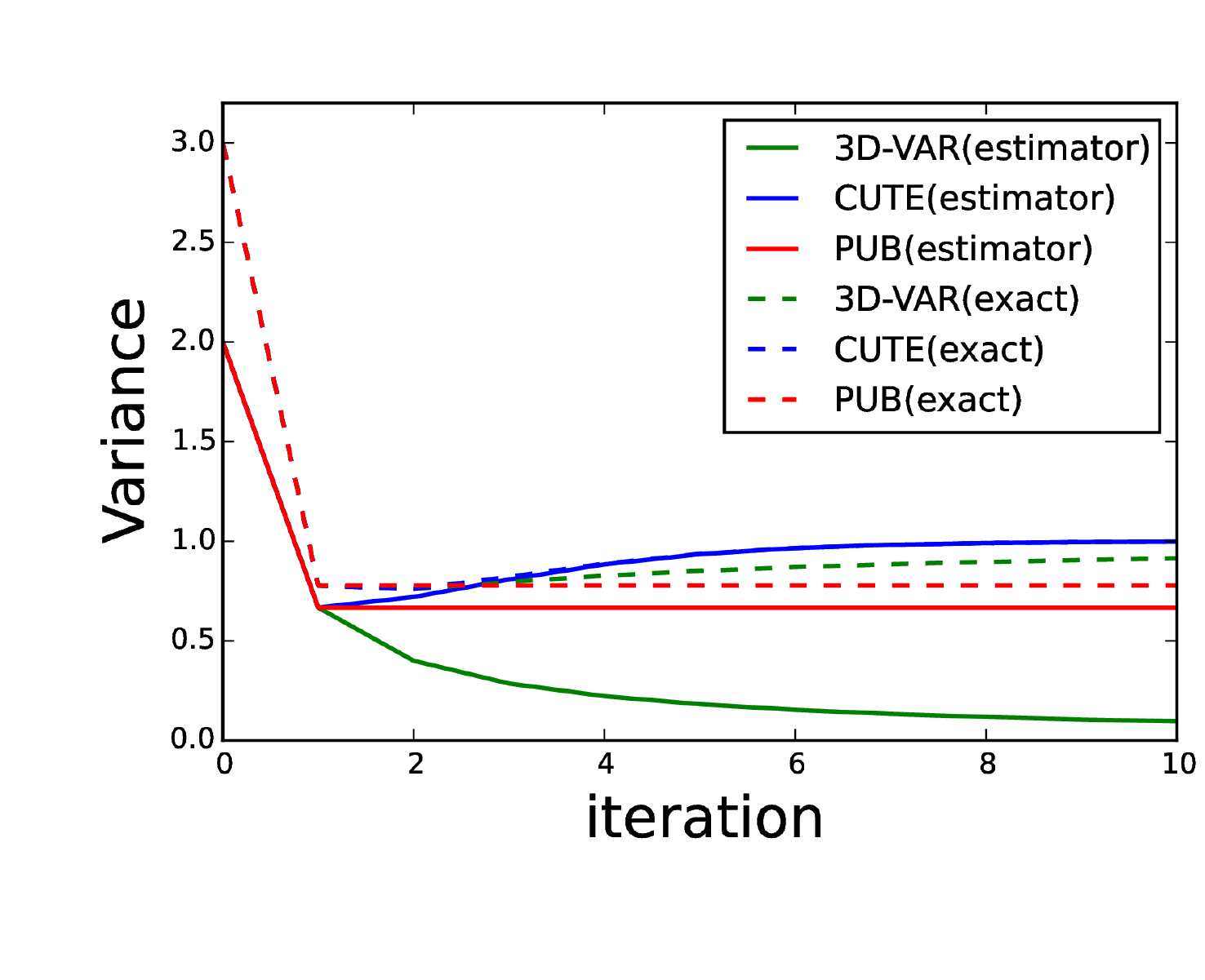}  
\caption{ Analysis of the evolution of the exact updated background error variance $\textbf{B}_n$ (dashed curves) vs its estimation provided by data assimilation algorithms $\textbf{B}_{\textsc{A},n}$ (solid curves). On the left side, the prior error variance is perfectly known (i.e. $\textbf{B}_{\textsc{A}}=\textbf{B}_\text{E}=3$) at the initial step.
On the right side, the background variance is voluntarily under-estimated:  ($\textbf{B}_{\textsc{A}}=2,\textbf{B}_\text{E}=3$). We remind that the updating of $\textbf{B}_{\textsc{A},n}$ is independent of the exact covariance evolution $\textbf{B}_{\text{E},i=1,\ldots,n}$; however, $\textbf{B}_{\text{E},n}$ is a function of the recurrence $\textbf{B}_{\textsc{A},i=1,\ldots,n-1}$.The observation error variance is  fixed at $R=1$, perfectly known for both solid and dashed lines.}
\label{fig:1d_analyse}
\end{figure}

Going back to the simple numerical illustration of a scalar case introduced in section \ref{sec:Iterative methods with advanced covariance updating} and depicted in Fig.  (\ref{fig:1d_analyse}), we monitor the behaviour of our iterative algorithms for ten steps. Here, we display the evolution of
the successive analysed covariance matrices (in fact, we simply look at variances due to the scalar variables). The dashed lines represent the evolution of exact error variance for the different methods (i.e. $\textbf{B}_{\textsc{E},n}^{\text{Naive}}$, $\textbf{B}_{\textsc{E},n}^{\textsc{\cute}}$, $\textbf{B}_{\textsc{E},n}^{\textsc{\pub}}$), that we are capable of computing thanks to our perfect knowledge of the exact prior background variance.  The solid lines represent their associated estimators (i.e. $\textbf{B}_{\textsc{A},n}^{\text{Naive}}$, $\textbf{B}_{\textsc{A},n}^{\textsc{\cute}}$, $\textbf{B}_{\textsc{A},n}^{\textsc{\pub}}$). In the left figure, all algorithms start from a perfect knowledge of the prior background variance (i.e. $\textbf{B}_{\textsc{E}}=\textbf{B}_\textsc{A}$). \rtwo{Since we are dealing with a scalar problem, there is no need to adjust the matrix trace, while $\alpha$ is set to be one in all applications of CUTE/PUB (i.e. $\textbf{B}_{\textrm{A},n+1}=\textbf{A}_{\textrm{A},n} $). }

In this case, the estimated variances provided by \cute and \pub coincide with the evolution of the exact variance. Meanwhile the estimated variance of the naive approach converges to zero, which leads to a significant under-estimation. We remind the reader that the first step of these three iterative methods are the same. In the right figure, we voluntarily under-estimate the exact background error variance at the beginning. We notice that  $\textbf{B}_{\textsc{A},n}^{\text{\cute}}$ and $\textbf{B}_{\textsc{A},n}^{\text{\pub}}$ are stable after some iterations. This behaviour was verified (not displayed here) no matter the choice of the initial variance. Moreover, for \cute method, we notice that the estimation of error variance becomes consistent with the exact error variance and they both converge to the observation error variance. \rone{Meanwhile despite being under-estimated by its estimator (solid red line), the exact error variance of PUB (dashed red line) remains inferior to the one of CUTE and \textit{3D-VAR} (dashed green line). } In both situations, a simple naive iteration of the variational method (green solid lines) leads to an important under-estimation (green curves) of the posterior error variance.

Unlike our illustration of the scalar case, in a space of larger dimension, these iterative methods may not reach a convergence in terms of error covariance and analysis state. We will discuss later how to define the stopping criteria outside the framework of twin experiments. \rone{Under the assumption of lower noise observation level}, one well-known quantity that has to be monitored is the innovation quantity: $(\textbf{y}-\textbf{H}(\textbf{x}_{b,n}))$ \rone{which will be displayed in the following numerical tests (e.g. Fig. \ref{fig:relative_exp_3}, Table. \ref{table:1})}.

\section{Numerical experiments} \label{sec:numeric}
\begin{figure}[!ht]
  \centering
    \includegraphics[trim=2cm 5cm 4cm 11cm,clip=true,width=4.5in]{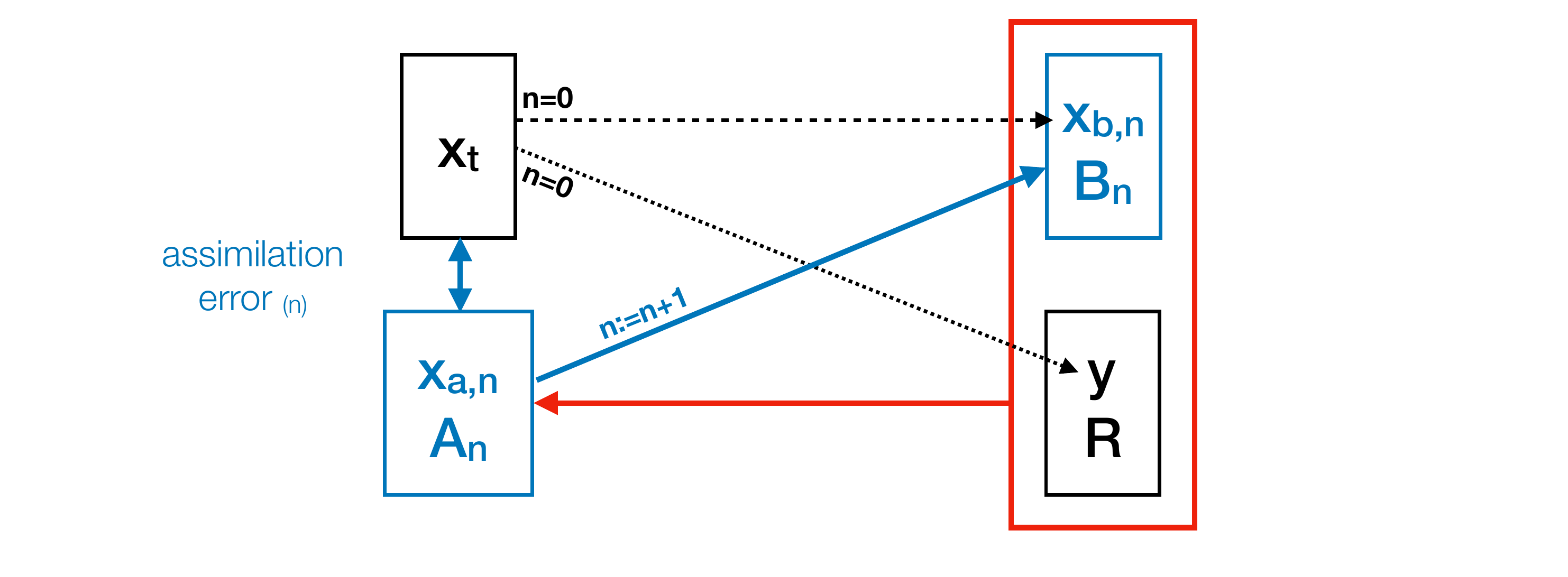}
\caption{Scheme of a twin experiments data assimilation framework for an iterative method.  Quantities in black are kept fixed while iterations are repeated: new assimilated state $\textbf{x}_{a,n}$ and covariance errors $\textbf{A}_{n}$ are injected  at the next step in order to update background quantities. The difference (in some norm) between the true state and the output of the algorithm $||\textbf{x}_t-\textbf{x}_{a,n}||$ is called the error of reconstruction and may be monitored. The entire experiment may be repeated numerous times for different realisations of $\textbf{x}_{b,n=0}$ to collect statistics of the assimilation results in order to assess the method robustness.}
\label{fig:twin_ex}
\end{figure}

Numerical experiments in twin experiments framework are carried out in order to compare the performance of the different methods. This principle is illustrated in Fig. \ref{fig:twin_ex} where the background states and the observations are obtained from a chosen true state by adding a {\em known} artificial noise (dashed line). The objective is to estimate how close is the estimated output to the true state. In this work, it is quantified by computing the expectation of the assimilation error $\mathbb{E}(||\textbf{x}_t-\textbf{x}_a||_{L^2})$ over the support of the {\em a priori} noise level, relying on Monte Carlo tests with different realizations of $(\textbf{x}_b, \textbf{y})$. More precisely, the original background state $\textbf{x}_{b,n=0}$ ($n$ stands for the number of iterations in CUTE, PUB) and observation $\textbf{y}$ (via $\textbf{H}$) are first constructed from a chosen true state $\textbf{x}_t$, thanks to the exact knowledge of $\textbf{B}$ and $\textbf{R}$. In our experiments, $\textbf{x}_t$ is obtained by a reference simulation. 

\subsection{Description of the system} 

\begin{SCfigure}[][!ht]
  \begin{minipage}{.4\textwidth}
\begin{align}
    \frac{\partial u}{\partial t}&=-g\frac{\partial}{\partial x}(h)-bu \label{eq: sw} \\
    \frac{\partial v}{\partial t}&=-g\frac{\partial}{\partial y}(h)-bv  \notag\\ 
    \frac{\partial h}{\partial t}&=-\frac{\partial}{ \partial x}(uh)-\frac{\partial}{\partial y}(v h)  \notag \\
    u_{t=0} &= 0 \notag \\
    v_{t=0} &= 0. \notag
\end{align}
  \end{minipage}
    \begin{minipage}[H]{.0\textwidth}
        \includegraphics[width=2.8in]{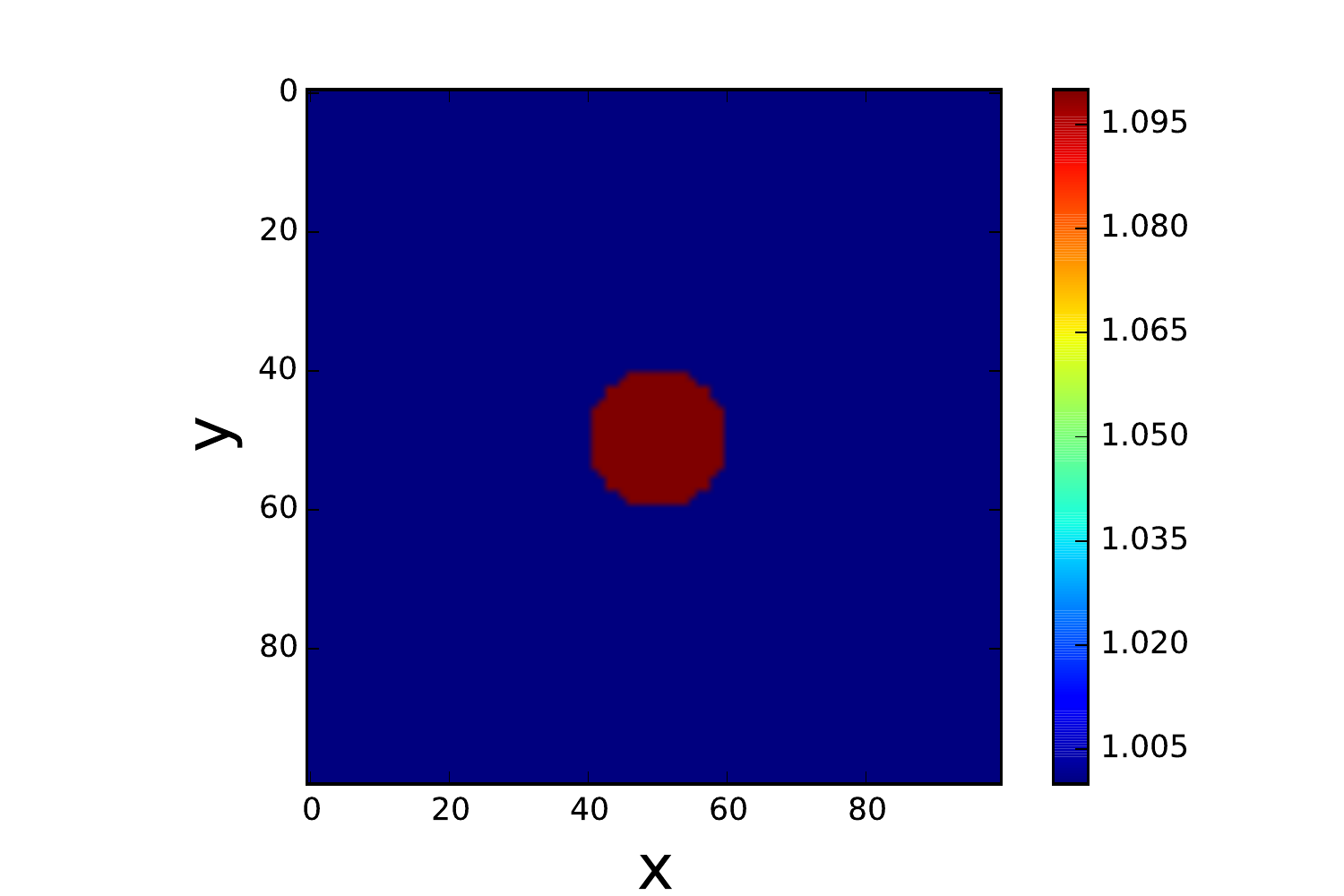}
        \\
        \\
        \caption{Initial $h$  of shallow water in $mm$}
        \label{fig:h_initial}
  \end{minipage}
\end{SCfigure}

In the following twin experiments, we consider a standard shallow-water fluid mechanics system which is frequently used for evaluating the performance of data assimilation algorithms (as in \cite{dance2013}, \cite{Cioaca2014}). The wave-propagation problem is nonlinear and time-dependent. The initial condition is chosen in the form of a cylinder of water of a certain radius that is released at  $\textit{t}=0$. We assume that the horizontal length scale is more important than the vertical one and we also neglect the Coriolis force. They lead to the  Saint-Venant equations (\cite{Venant1871}) coupling the fluid velocity and height as shown in Eq.\ref{eq: sw} where $(u,v)$ are the two components of the two dimensional fluid velocity (in $0.1m/s$) and $h$ stands for the fluid height (in millimeter). The earth gravity constant $g$ is thus scaled to 1 and the dynamical system is defined in a non-conservative form.

\begin{figure}[!ht]
  \centering
      \begin{subfigure}[H]{.23\textwidth}
     \includegraphics[trim=3cm 8cm 3cm 9.5cm,clip=true,width=1.7in]{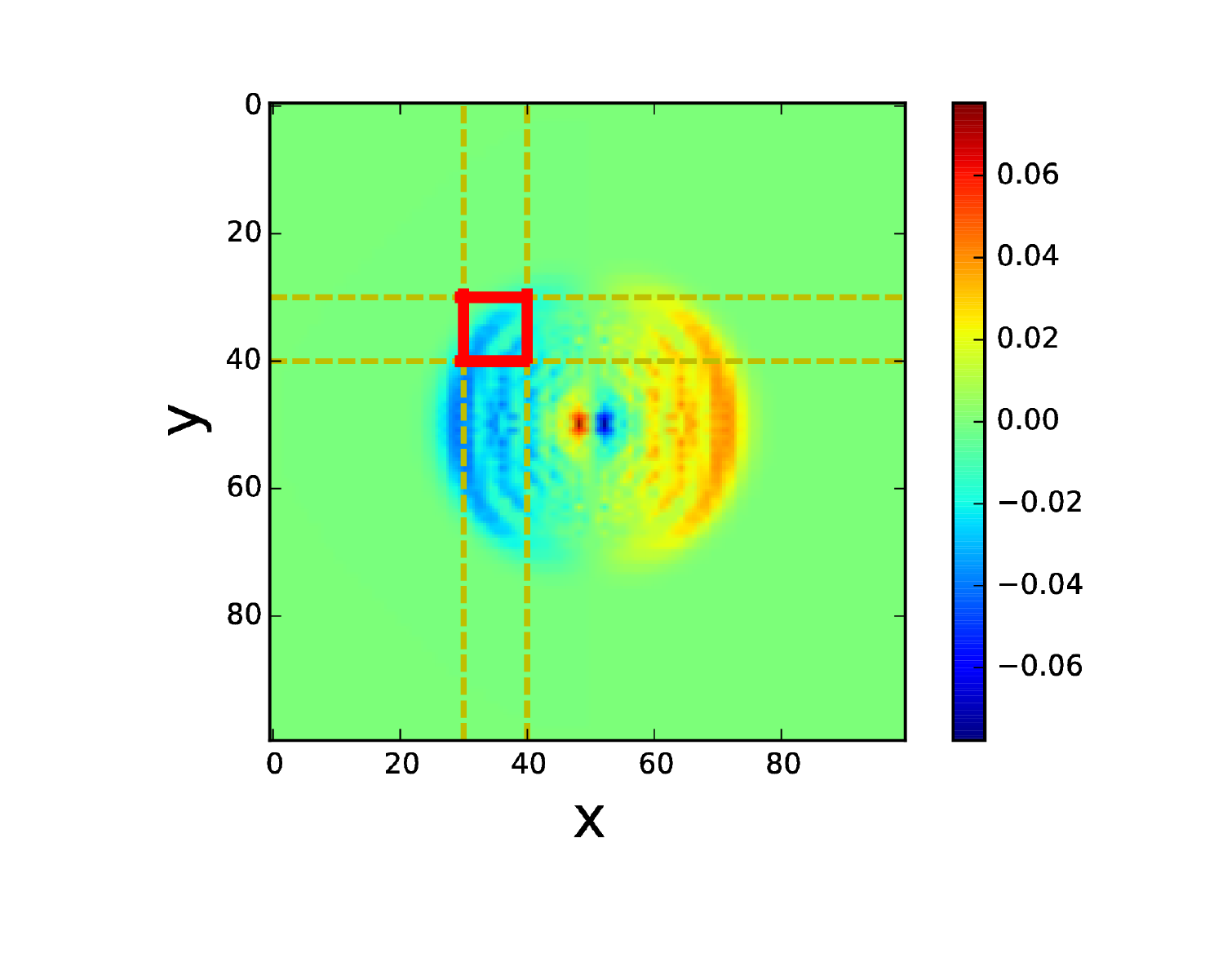}
      \subcaption{}
    \end{subfigure}    
    \begin{subfigure}[H]{.2\textwidth}
     \includegraphics[trim=3cm 8cm 3cm 9.5cm,clip=true,width=1.7in]{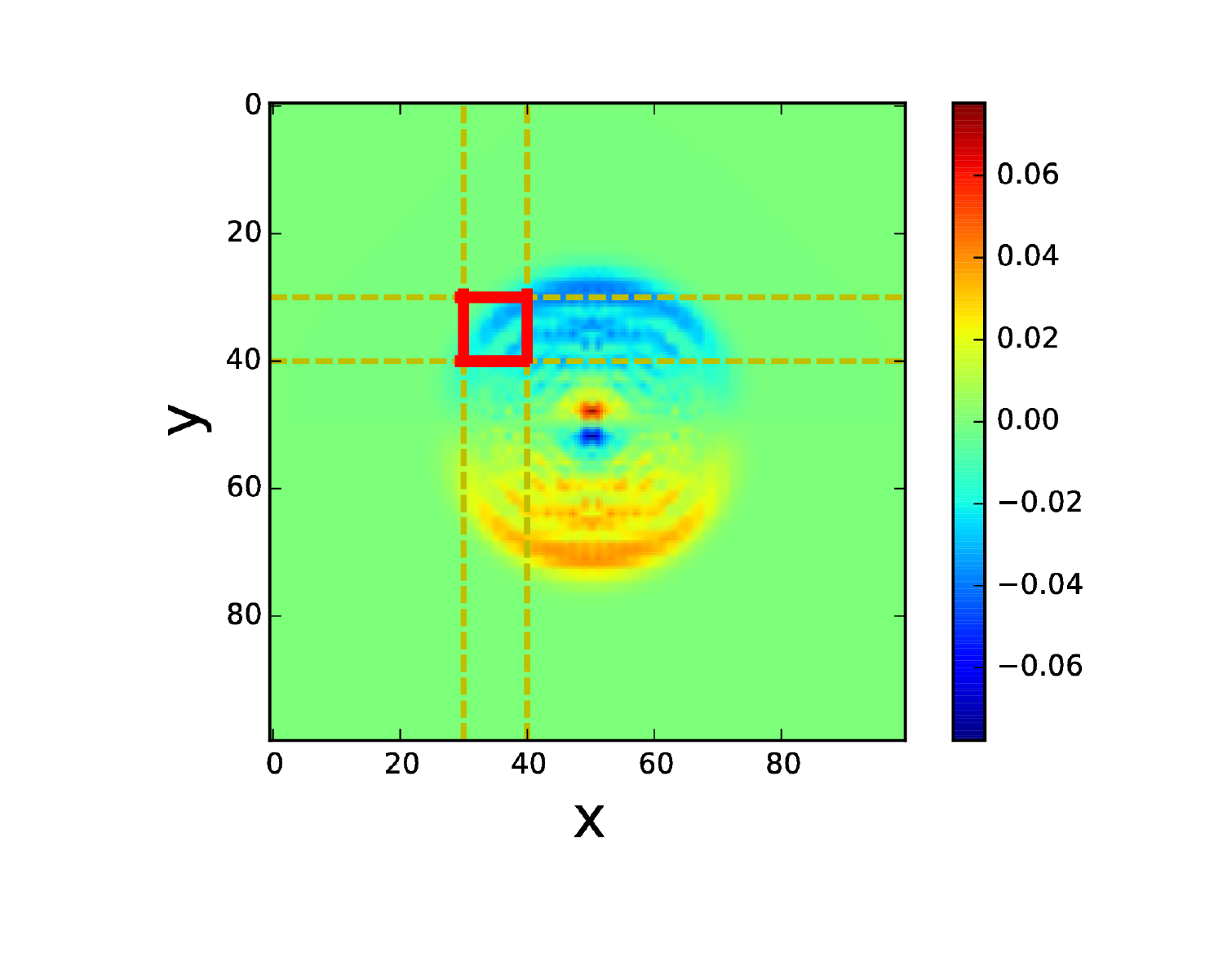}
      \subcaption{}
    \end{subfigure}
     \begin{subfigure}[H]{.26\textwidth}
      \includegraphics[trim=2cm 8cm 3cm 9.5cm,clip=true,width=2.2in]{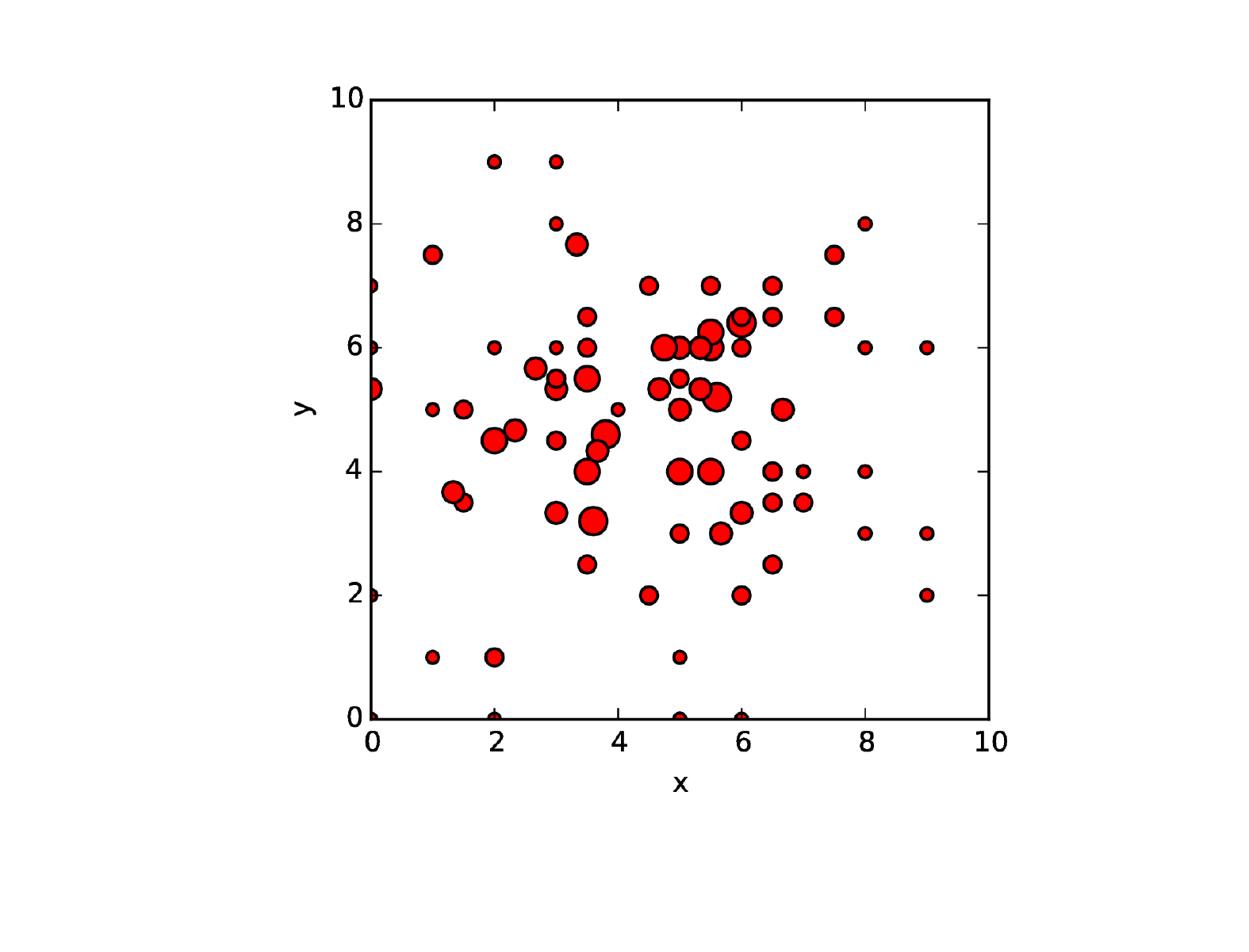}
      \subcaption{}
    \end{subfigure}
    \begin{subfigure}[H]{.25\textwidth}
      \includegraphics[trim=2cm 8cm 4cm 9.5cm,clip=true,width=1.7in]{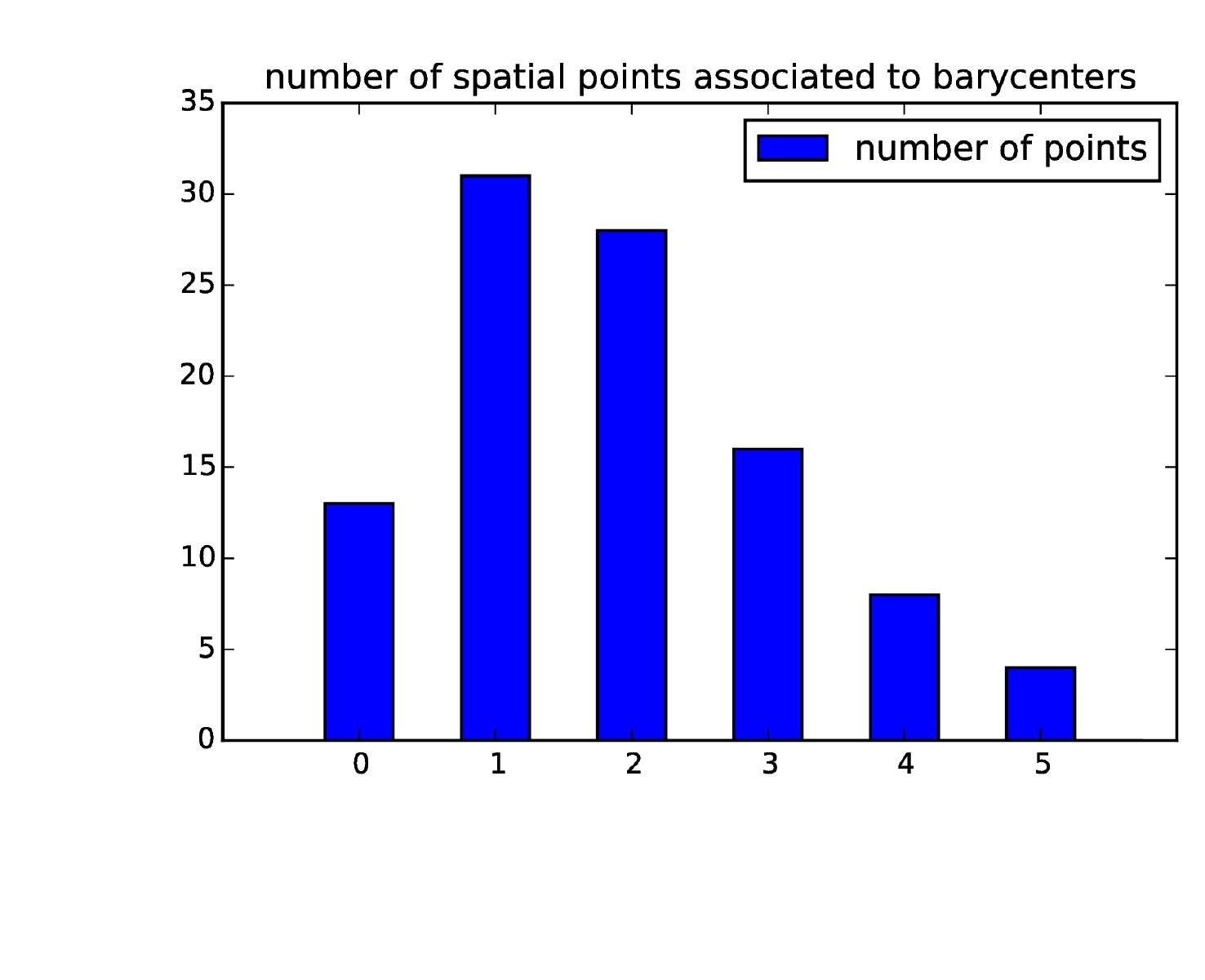}
      \subcaption{}
    \end{subfigure}

\caption{Illustrations of the random observation operator $\textbf{H}$ and example of 2D flow velocity fields of the shallow water model.  Barycenters measured by the linear transformation operator $\textbf{H}$ are shown in (c) where the symbol radius is proportional to the number of measures associated to each barycenter. The histogram of the number of selected points associated to each barycenter (rows in matrix $\textbf{H}$) is shown in (d) and is reminiscent of a binomial distribution. Shallow water 2D velocity fields are represented in (a) and (b) (respectively for $u$ and $v$ in Eq. (\ref{eq: sw}))  at time $t=t_f$. Data assimilations are performed in the red square subdomain.}
\label{fig:H with braycenter}
\end{figure}

\noindent  The initial values of $u$ and $v$ are set to zero for the whole velocity field and the height of the water cylinder is set to be $h^{\textsc{cyl}}_{t=0}=0.1mm$ high above the one of the still water as shown in Fig. \ref{fig:h_initial}. The domain of size $(L_x\times L_y)=(100mm \times 100mm)$ is discretized with a regular structured grid of size $(100 \times 100)$ and the solution of Eq. (\ref{eq: sw}) is approximated thanks to a finite difference method of first order. The time-integration is also first-order with a time interval $\delta t=10^{-6} s$. The system is integrated up to a time $t_f = 1.5 \times 10^{-3} s$ (see Fig. \ref{fig:H with braycenter}(c-d)), and the obtained solution is used as the reference state ($\textbf{x}_t$) in \ref{subsec:homo_error} and \ref{subsec:relative_error}. Our objective  is to reconstruct the state $x=(u,v)$ within a non-centered $(10 \times 10)$ subdomain (represented by a red square in Fig. \ref{fig:H with braycenter} (c) and (d)) from noisy measurements via data assimilation processes. Thanks to an observation operator $\textbf{H}$ described later, we will use a collection of observations from the subdomain. Therefore, the dimension of the state space (i.e. $\textbf{x}_t, \textbf{x}_a, \textbf{x}_b$) which combined two 2D fields $u$ and $v$ may be algebraically combined in an array of size 200, i.e. $\textsc{x}_t \equiv \left \{ \textsc{x}_t(k) \right \}_{k=1\ldots 200}\equiv \left\{ (u (t=t_f), v (t=t_f)) \right\} $. The observation vector $\textbf{y}$ is of size $100$ but with zero elements included as shown in Fig. \ref{fig:H with braycenter} (b).\\

In all numerical tests, we keep the assumption of linearity of the transformation operator $\mathcal{H}$. As we have mentioned in section \ref{sec:Introduction to 3D-Var}, $\mathcal{H}$ could represent a transformation between two physical quantities/fields or even include discretized forecast/model operators. In this work, we wish to remain as general as possible. Therefore, we prefer not to set a particular form of the observation operator, which would promote some space-filling properties or some other type of optimality. 
With this aim, we decide to model the observation operator with  a random matrix $\textbf{H}$ acting as a binomial selection operator.
Each observation will be constructed as a sum of a contribution from a linear combination of a few true state variables randomly collected over the subdomain and some random noise.
In order to do so, 
we introduce the notation for a subset sample $\left \{ \textsc{x}_t^*(i) \right \}_{i=1\ldots n}$ randomly but homogeneously chosen (with replacement) with probability $P$ among the available data set, i.e. $\left \{ \textsc{x}_t(k) \right \}_{k=1\ldots 200}$. The subset values $\textsc{x}_t^*$ are summed up and the process is re-iterated 100 times in order to construct the observations:
\begin{equation}
    y(j) = \sum_{i=1}^{n_j}\textsc{x}_t^*(i) + \epsilon_y, \quad \text{for} \; j=1,\ldots,100,
\end{equation}
where the size $n_j$ of the collected sample used for each $j^{\text{th}}$ observation data point $y(j)$ is random and by construction follows a binomial distribution $\mathcal{B}(200,P)$. In the following we choose a sparse representation with $P=1\%$.
~\\

Once $\textbf{H}$ is randomly chosen, it is kept fixed for a whole set of numerical experiments. This operator $\textbf{H}$ is shown in Fig. \ref{fig:H with braycenter} ((c) and (d)). In fact, with this definition of $\textbf{H}$, the  observed quantities can be apprehended as some sorts of barycenters in the state space. As explained, the number of points in the field associated to each barycenter can thus be seen as a random variable of binomial distribution as shown in Fig. \ref{fig:H with braycenter} (d). If we increase the probability of success of the selection operator, more points will be selected and combined across the domain, resulting in a more centered barycenters distribution. 

\rtwo{We have numerically verified that in general, the results obtained with a transformation operator $\textbf{H}$ of more regular span structure tend to be less optimal than the ones obtained in the case with randomly simulated $\textbf{H}$, in terms of output correlation identification. We believe that repeated assimilations based on the same uniform data set of observations may unwillingly put emphasis on certain correlation lengths while missing others, in relation to the structure of $\textbf{H}$.}

\subsection{Experiments with state-independent homogeneous prior errors} \label{subsec:homo_error}
For the sake of simplicity, in the analysis of data assimilation algorithms, \rone{prior background and observation errors (i.e. $\epsilon_b$ and $\epsilon_y$) are often supposed to be independent of their theoretical values (i.e. respectively $\textbf{x}_t$ and $\textbf{H} \textbf{x}_t$).} Under this assumption, the assimilation error depends only on the prior errors $\epsilon_b$ and $\epsilon_y$, as:
\begin{align}
    \textbf{x}_a-\textbf{x}_t&=\textbf{x}_b+\textbf{K}(\textbf{y}-\textbf{H} \textbf{x}_b)-\textbf{x}_t\\ \label{eq:homo1}
    &=\epsilon_b+\textbf{K}(\epsilon_y-\textbf{H} \epsilon_b) \notag\\ 
    &=(\textbf{I}-\textbf{K}\textbf{H}) \epsilon_b+\textbf{K} \epsilon_y. \notag
\end{align}
Therefore, the numerical results shown in this section are independent from the choice of the true state, and therefore valid for any (2D) field reconstruction with state-independent prior noise. Here, background states and observations are simulated using chosen error covariances matrices $\textbf{B}_{\textrm{E}}$ and $\textbf{R}$. Our assumption of higher background error amplitude leads to:
\begin{align}
    Tr(\textbf{B}_{\textrm{E}})>Tr(\textbf{R}).
\end{align}
In our experiments, the average standard deviation of the background error is set to be at least 10 times higher than the observation error.  We make further assumption that the correlation pattern of background covariance is poorly known. In order to make numerical tests representative, we make use of homogeneous and isotropic (invariant under rotations and translations) one-dimensional correlation patterns (of spatial euclidean distance $r= \sqrt{\Delta_x^2+\Delta_y^2}$) for simulating true or initially estimated background errors (i.e. $\textbf{B}_{\textrm{E}}$ and $\textbf{B}_{\textrm{A},n=0}$). We consider the following correlation function types:

\begin{itemize}
\item Exponential type: $\phi(r)=\exp(-\frac{r}{L})$, \label{eq:exp}
\item Balgovind type: $\phi(r)=(1+\frac{r}{L}) \exp(-\frac{r}{L})$, \label{eq:balgo}
\item Gaussian type:  $\phi(r)=\exp(-\frac{r^2}{2L^2})$,  \label{eq:Gaussian}
\end{itemize}
where $L$ is defined as the typical correlation length scale. These correlation functions are part of the Matérn family of covariance function (respectively of order $\nu = 1/2, 3/2$ and $\infty$) and are often used as imposed structures in background matrix construction (see \cite{Singh2011}, \cite{Poncot2013}). For the sake of simplicity, in this section  the correlation kernel of the exact background covariance matrices are always chosen to be of Balgovind type with scale length $L=2$, where observation errors are supposed to be spatially independent (i.e. $\textbf{R}$ is proportional to an identity matrix). The latter is supposed to be known in the algorithms. \rtwo{Because both the amplitude and the correlation pattern of $\textbf{B}_\textrm{E}$ are supposed to be poorly specified by $\textbf{B}_\textrm{A}$, we choose to set the coefficient of confidence $\alpha = 0$ for the trace operator in all following numerical tests.}\\

In order to verify the robustness of the proposed  methods, different scenarios are considered for the correlation pattern of the initial assumed covariance  $\textbf{B}_{\textsc{A},n=0}$. As mentioned in section \ref{sec:Iterative methods with advanced covariance updating}, the objective of our  algorithms is to improve the output error correlation estimation, and in consequence, obtain a reduction of assimilation error. In fact, using  Eq. (\ref{eq:correlation_covariance_2}), the error correlation matrices associated to $\textbf{B}_{\textsc{E},n}$ (exact background error correlation at $n^{\textrm{th}}$ step) and $\textbf{B}_{\textsc{A},n}$ (estimation of error correlation at $n^{\textrm{th}}$ step) can be extracted and compared at each iteration. Our objective is therefore to reduce the dissimilarity between these two correlation matrices, the monitoring of this distance taking different forms: -- through a simple correlation calibration or -- other correlation dissimilarity measure such as the \textit{Affine  Invariant  Riemannian  Metric}  (AIRM) (\cite{Cherian2011}) defined for two semi-positive definite matrices $\textbf{X}$ and $\textbf{Y}$ by:
\begin{align}
    D_{\textrm{AIRM}}(\textbf{X},\textbf{Y})=|| \log(\textbf{X}^{-1/2} \textbf{Y} \textbf{X}^{-1/2}) ||_F
\end{align}
where $||.||_F$ represent the Frobenius norm of matrices. This similarity measure is widely used as it integrates the knowledge of the manifold structure of the covariance matrices. In our cases, the two semi-positive definite matrices to be compared by AIRM are the assumed background error correlation matrix $\textbf{Cor}_{\textbf{B}_{\textsc{A},n}}$ and the exact background error correlation matrix $\textbf{Cor}_{\textbf{B}_{\textsc{E},n}}$ defined respectively from covariance matrices  $\textbf{B}_{\textsc{A},n}$ and $\textbf{B}_{\textsc{E},n}$ as shown in Eq. (\ref{eq:correlation_covariance_2}).
According to \cite{Pennec2006}, invariant under linear transformations, AIRM can be seen as a natural choice of metric for symmetric semi-positive definite matrices. \\

\begin{figure}[!ht]
  \centering
     \begin{subfigure}[H]{.3\textwidth}
      \includegraphics[trim=2cm 8cm 2cm 8.8cm,clip=true,width=2.2in]{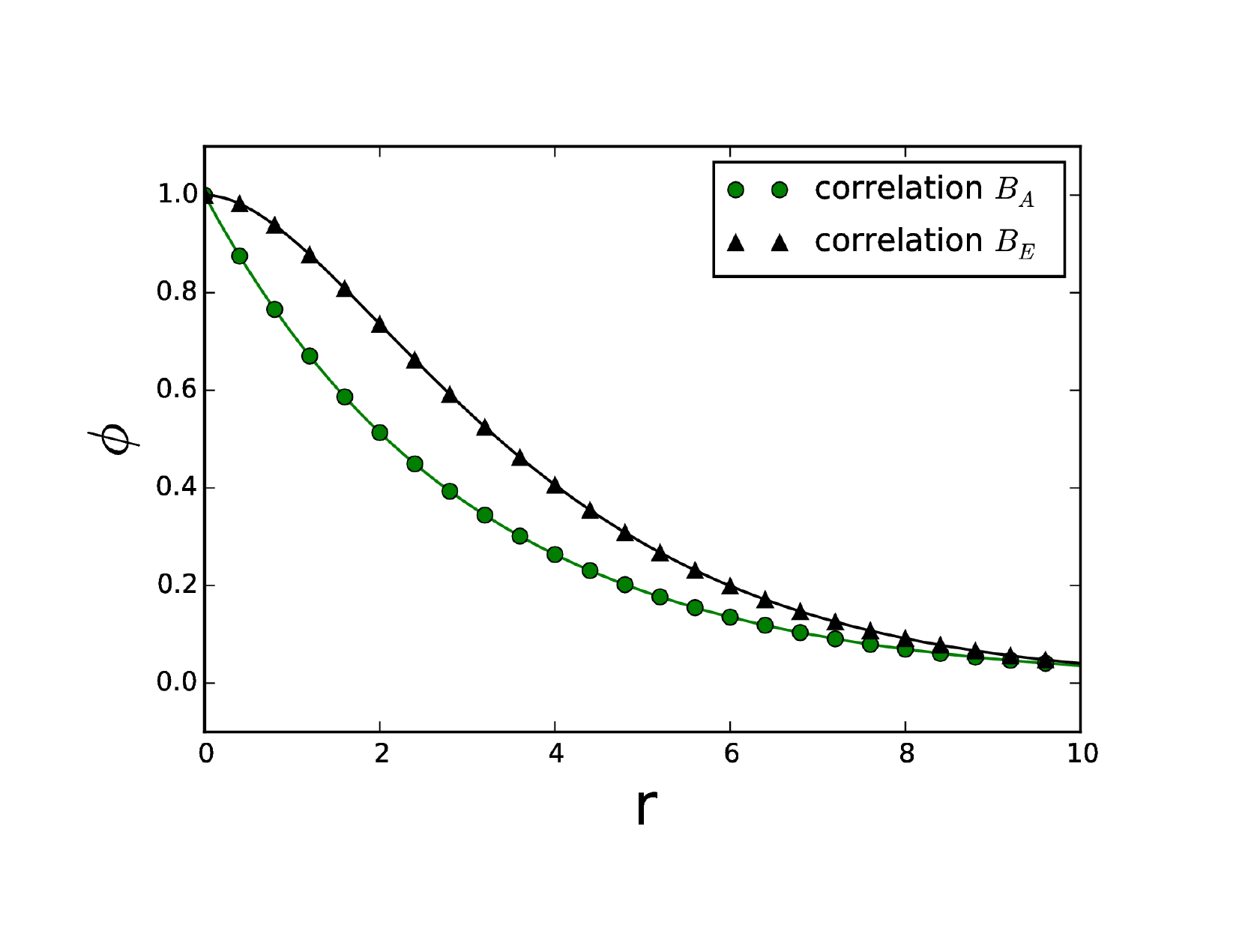}
      \subcaption{}
    \end{subfigure}
    \begin{minipage}[c]{.3\textwidth}
      \includegraphics[width=2.in]{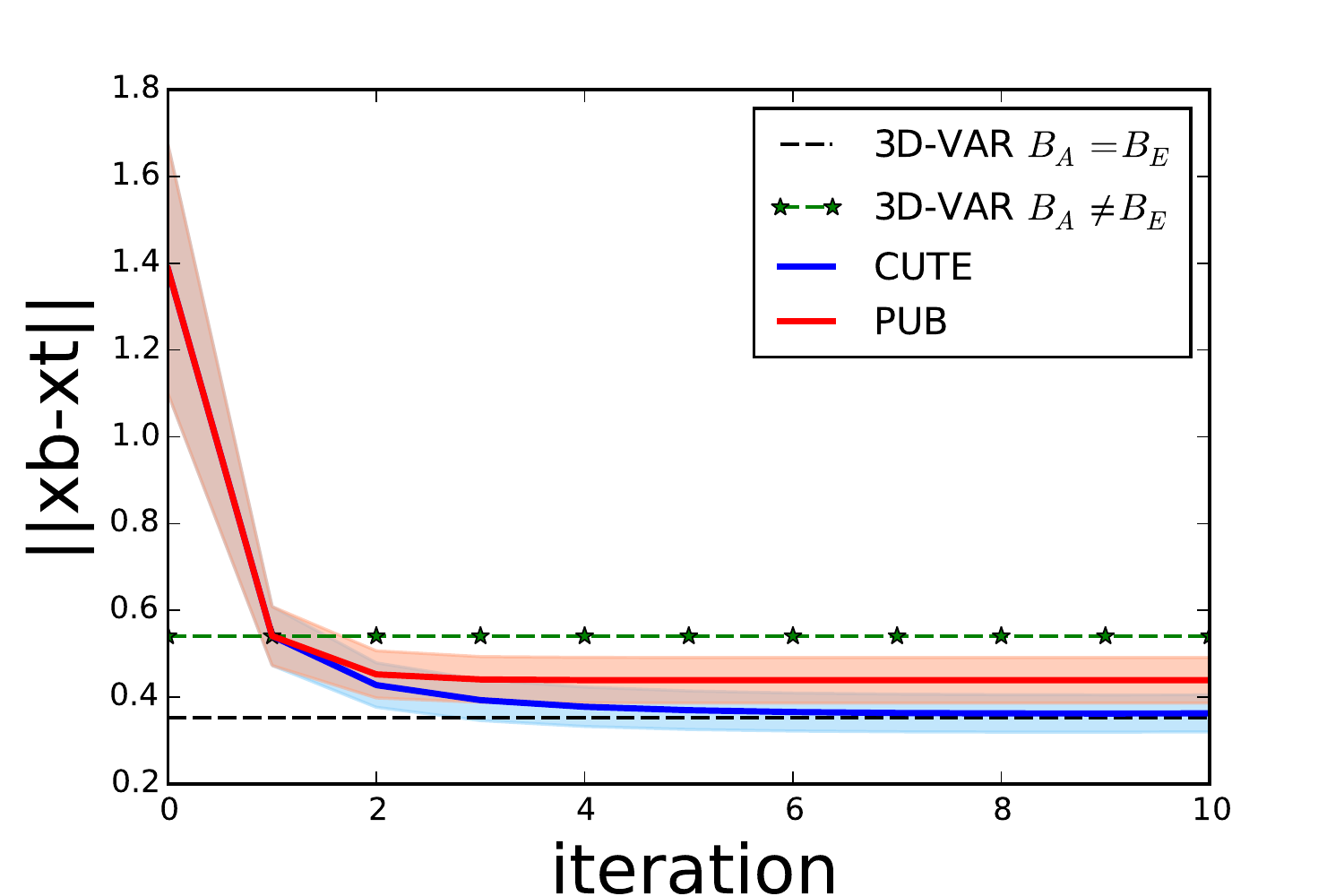}
      \subcaption{}
   \end{minipage}
    \begin{subfigure}[H]{.3\textwidth}
      \includegraphics[trim=2cm 8cm 2cm 8.8cm,clip=true,width=2.2in]{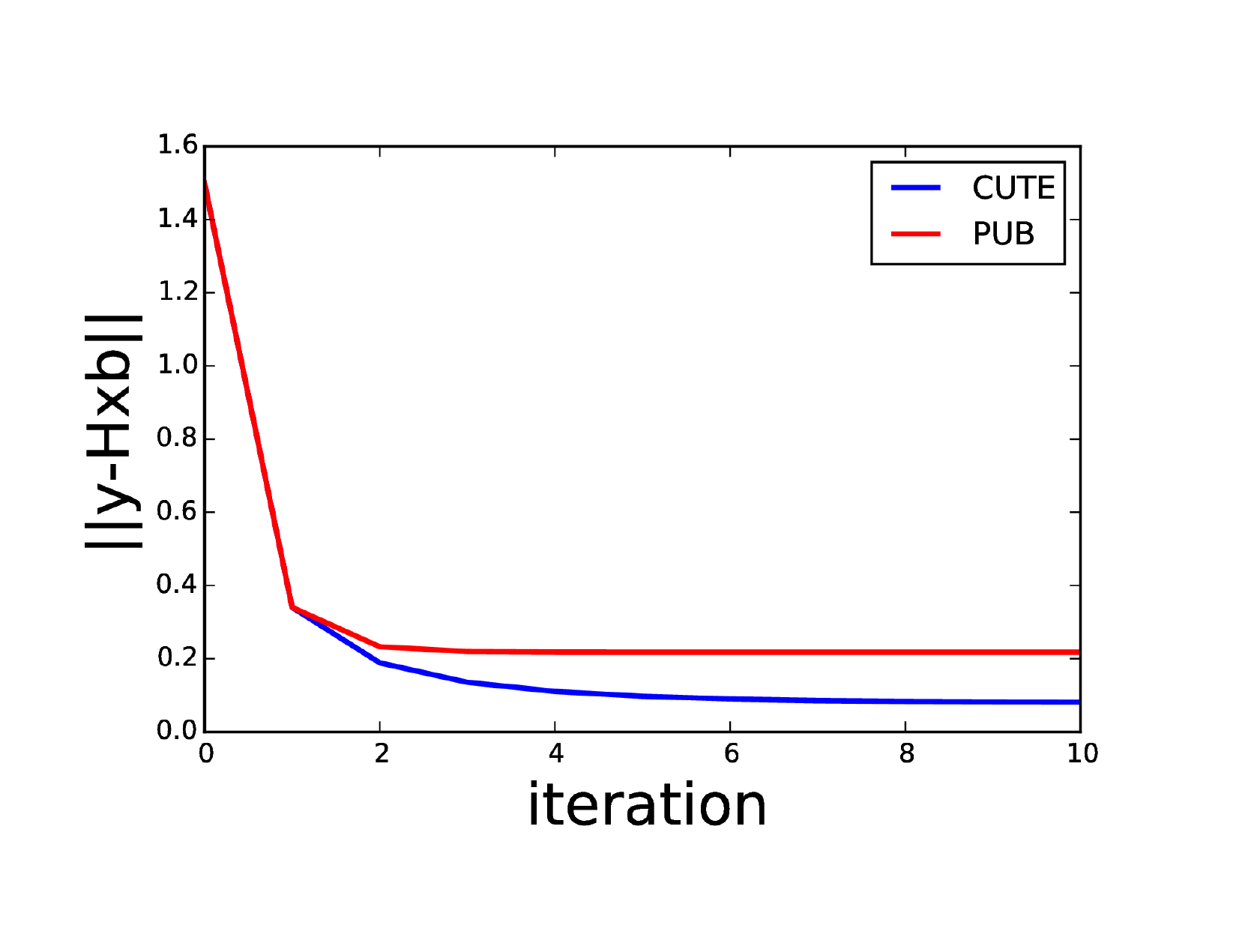}
      \subcaption{}
    \end{subfigure}        
    \begin{subfigure}[H]{.3\textwidth}
      \includegraphics[trim=2cm 8cm 2cm 8.8cm,clip=true,width=2.2in]{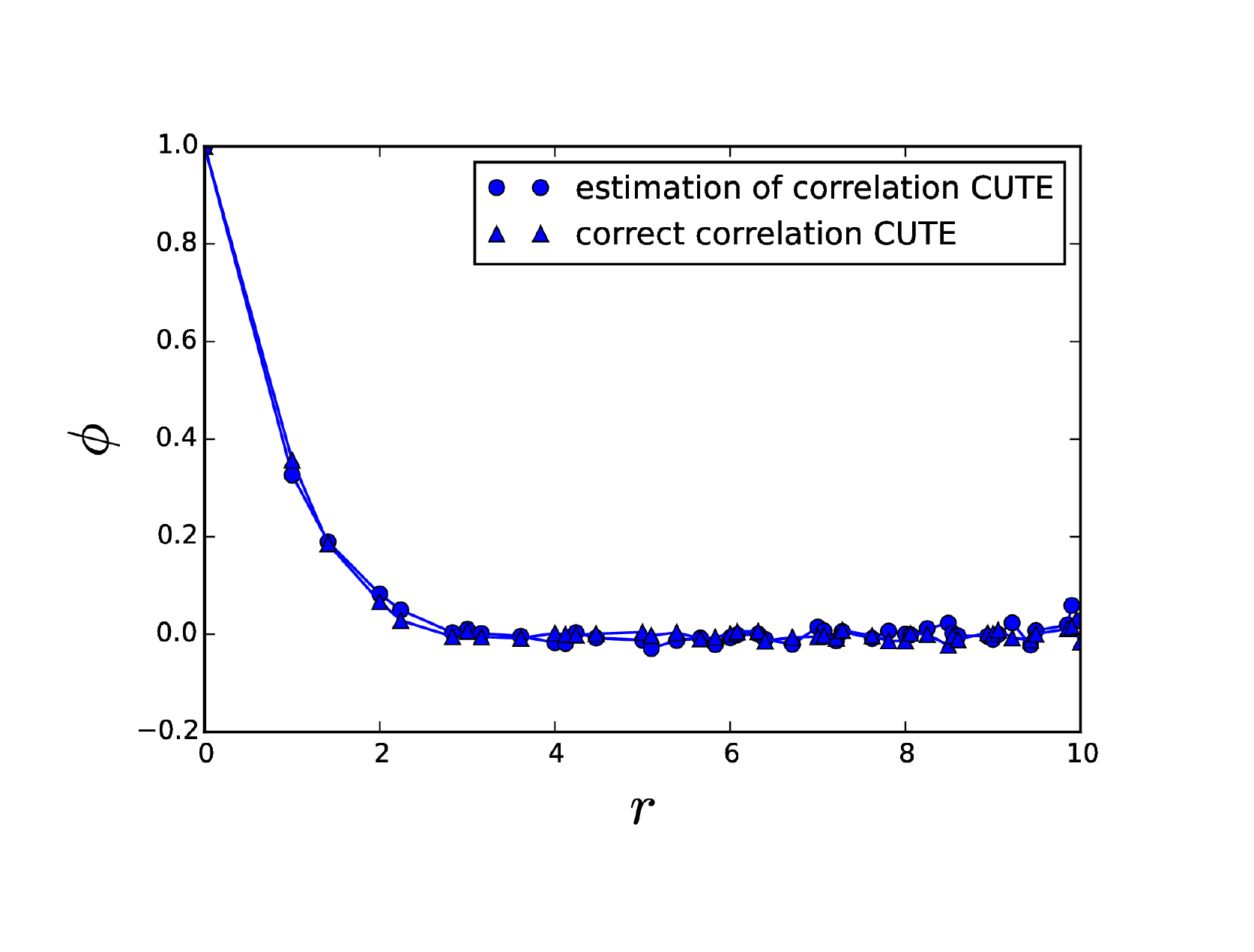}
      \subcaption{}
    \end{subfigure}
    \begin{subfigure}[H]{.3\textwidth}
      \includegraphics[trim=2cm 8cm 2cm 8.8cm,clip=true,width=2.2in]{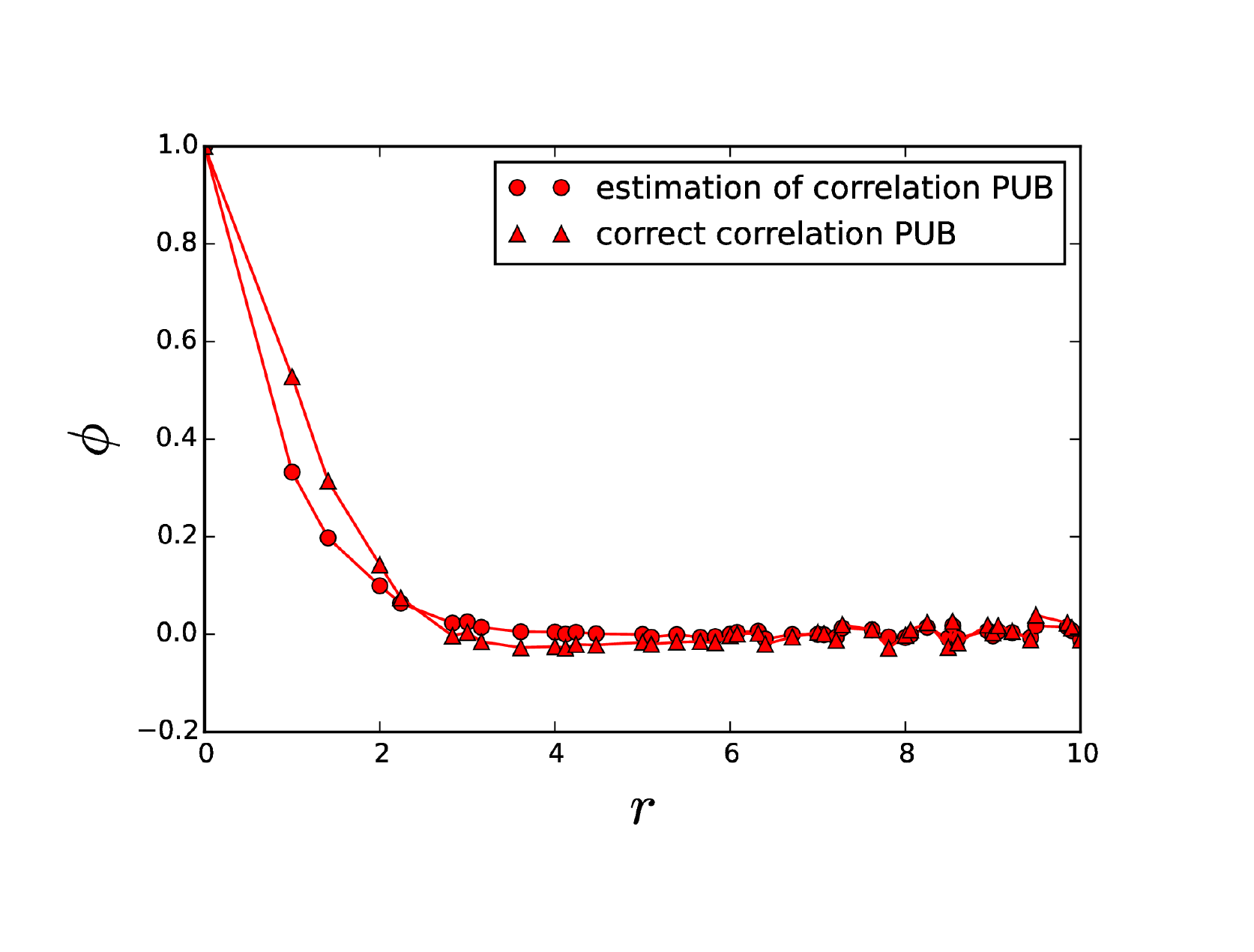}
      \subcaption{}
    \end{subfigure}
    \begin{subfigure}[H]{.3\textwidth}
      \includegraphics[trim=2cm 8cm 2cm 8.8cm,clip=true,width=2.2in]{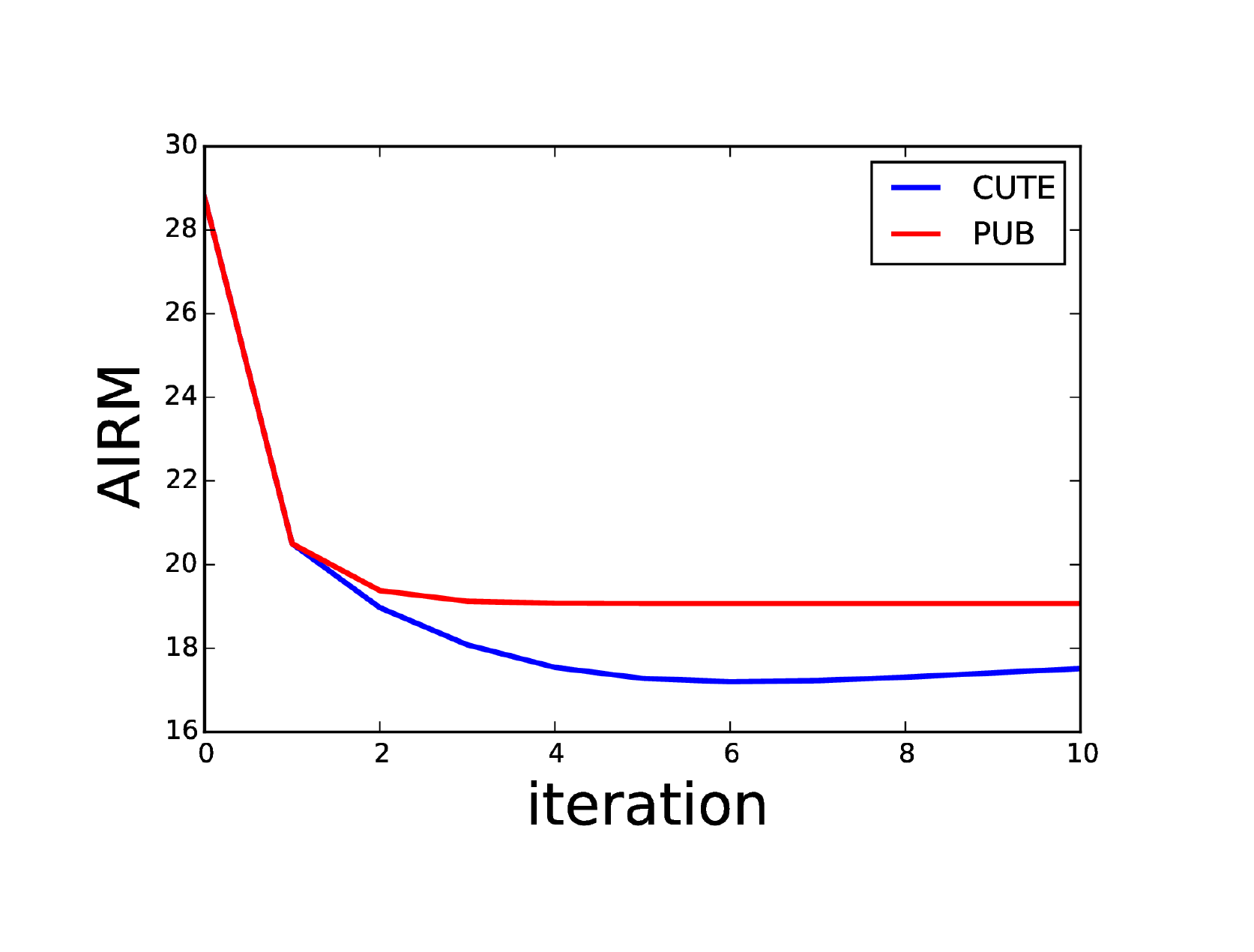}
      \subcaption{}
    \end{subfigure}
\caption{ Twin experiments with state-independent homogeneous prior error. Figures on the first line refer to the initial choice of prior correlations (a) and the evolution of assimilation error (b) and innovation quantities (c), while figures on the second line monitor iterated quantities extracted from the errors covariance in the velocity field of $u$ (d-f).  In this test, $\textbf{B}_{\textsc{A},n=0}$ is chosen to follow an exponential kernel with $L=3$ (shown by the green curve in (a)).
}
\label{fig:homo_exp_3}
\end{figure}

\begin{figure}[!ht]
  \centering
     \begin{subfigure}[H]{.3\textwidth}
      \includegraphics[trim=2cm 8cm 2cm 8.8cm,clip=true,width=2.2in]{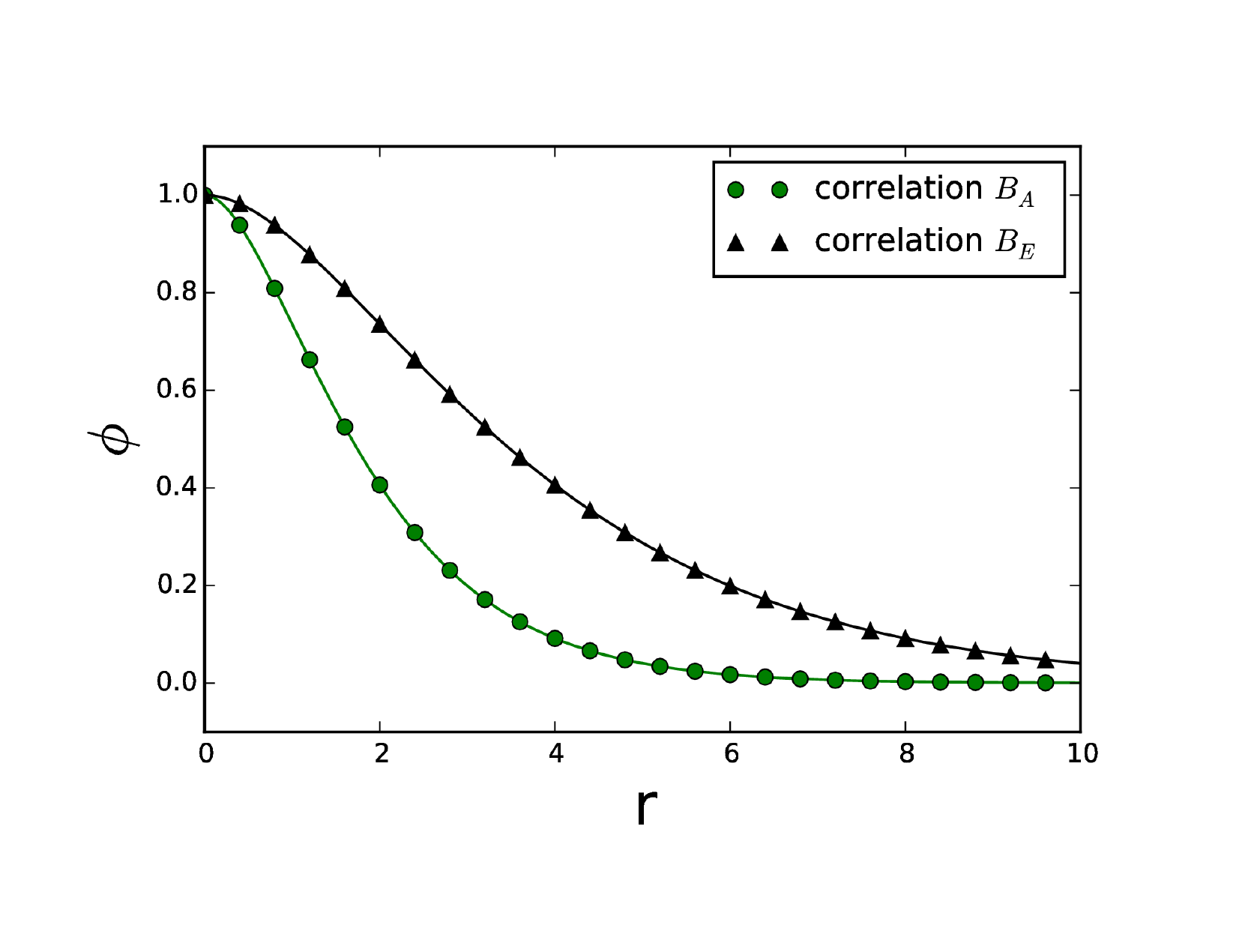}
      \subcaption{}
    \end{subfigure}
    \begin{subfigure}[H]{.3\textwidth}
      \includegraphics[trim=2cm 8cm 2cm 8.8cm,clip=true,width=2.2in]{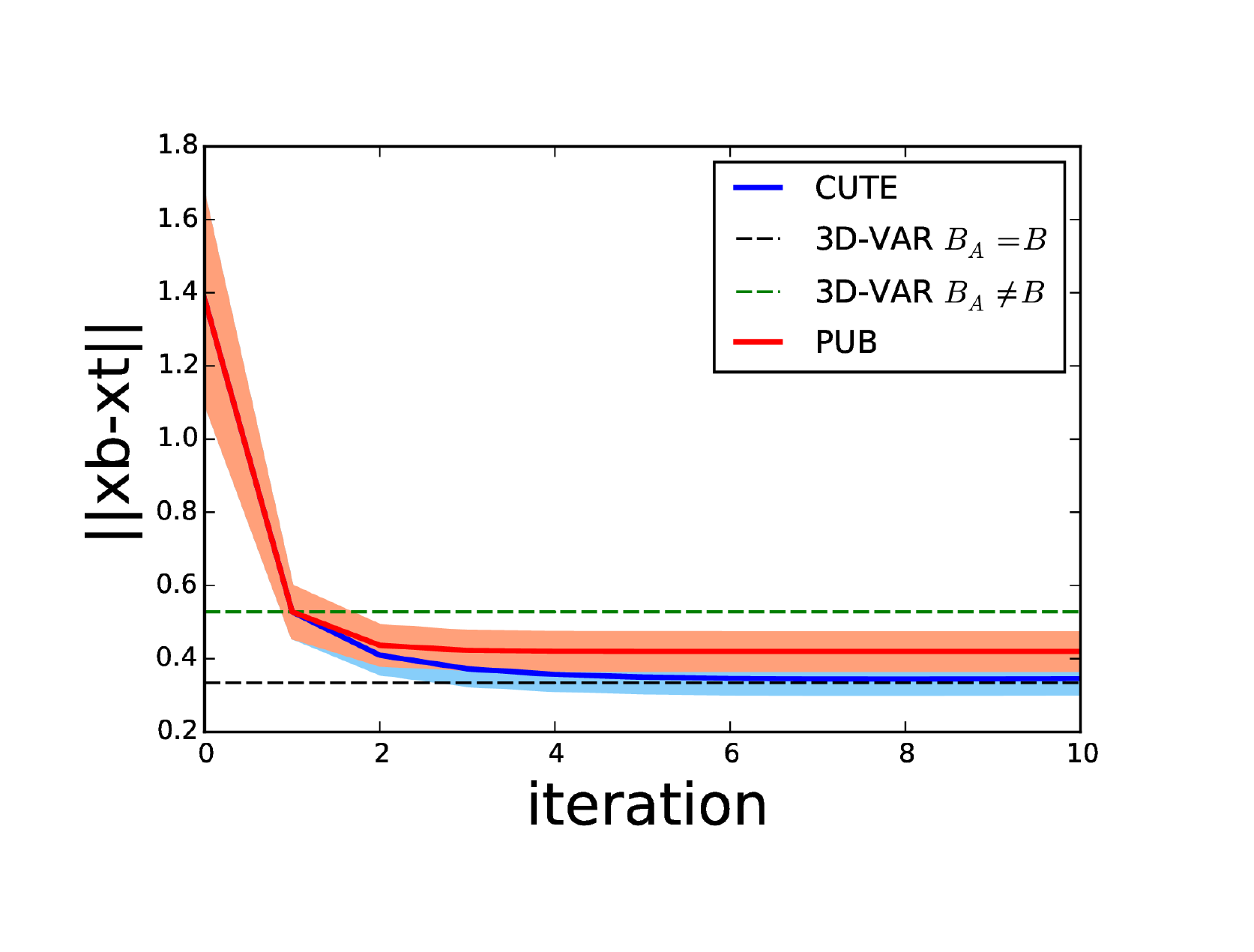}
      \subcaption{}
   \end{subfigure}
\caption{Evolution of assimilation error in twin experiments using same simulated observations as Fig. \ref{fig:homo_exp_3} with different initial background matrix estimation ($\textbf{B}_{\textsc{A},n=0}$ is of Balgovind type with scale length $L=1$).}
\label{fig:homo_bal_1}
\end{figure}

\begin{figure}[!ht]
  \centering
     \begin{subfigure}[H]{.3\textwidth}
      \includegraphics[trim=2cm 8cm 2cm 8.8cm,clip=true,width=2.2in]{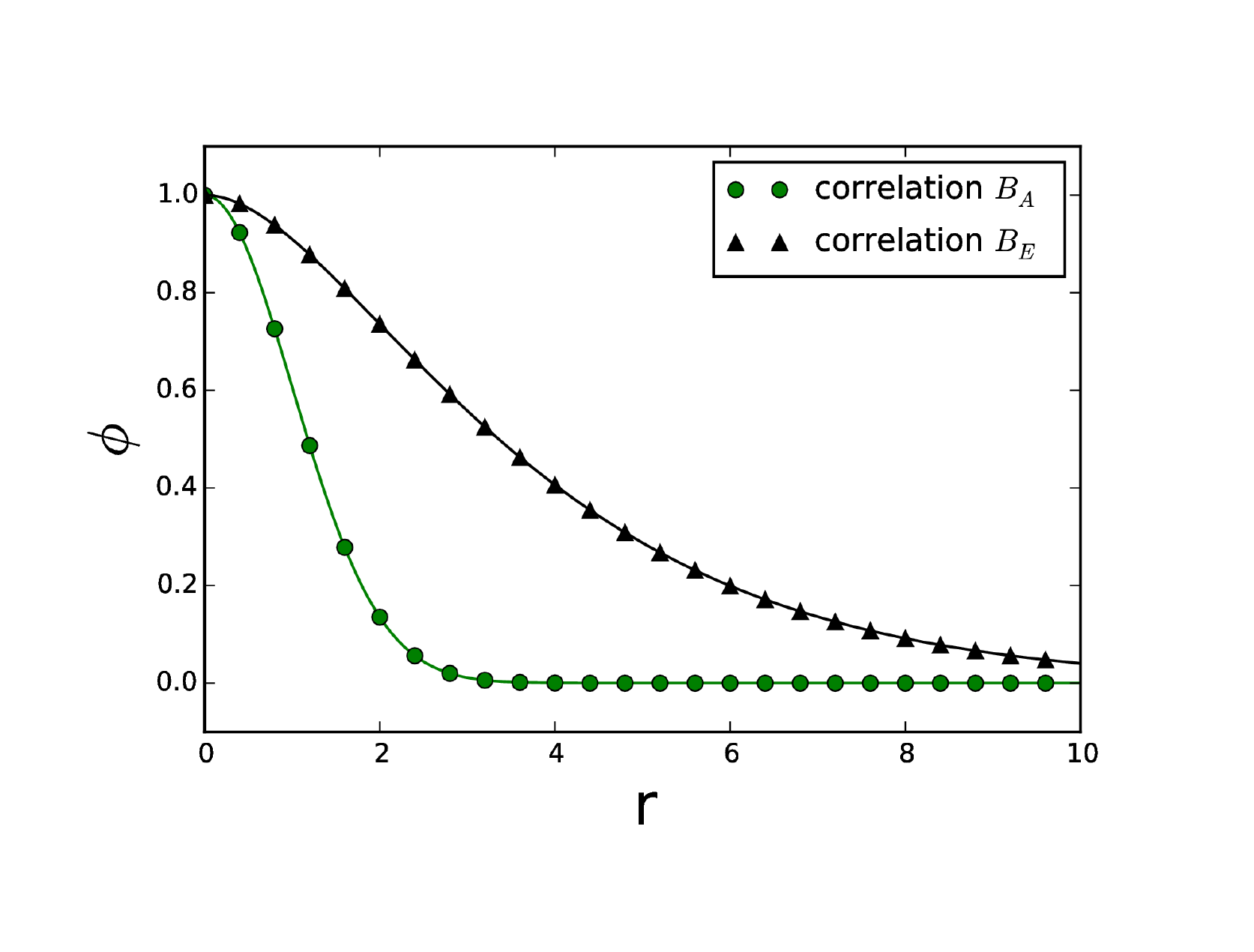}
      \subcaption{}
    \end{subfigure}
    \begin{subfigure}[H]{.3\textwidth}
      \includegraphics[width=2.in]{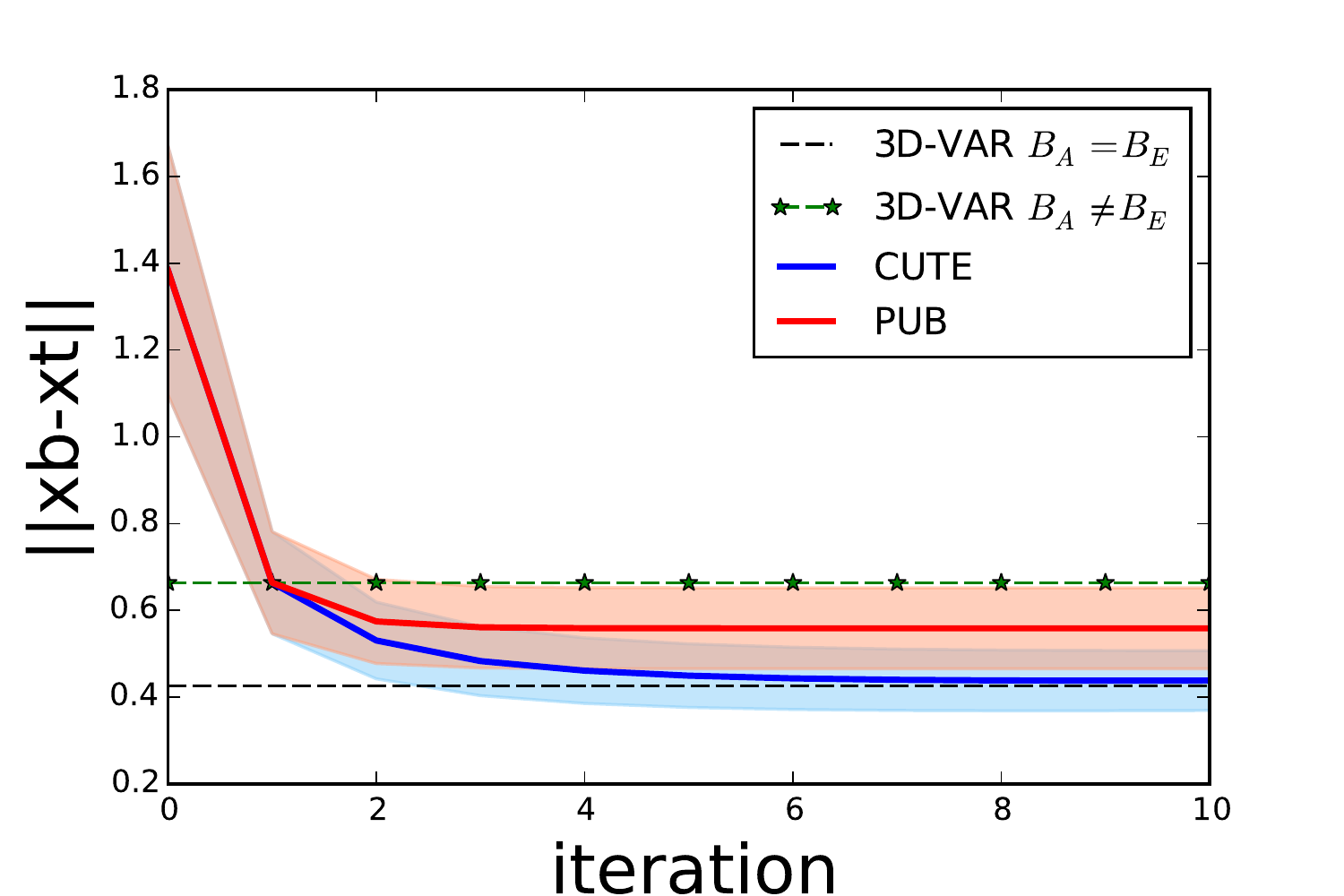}
      \subcaption{}
   \end{subfigure}

\caption{Evolution of assimilation error in twin experiments using same simulated observations as Fig. \ref{fig:homo_exp_3}  with different initial background matrix estimation ($\textbf{B}_{\textsc{A},n=0}$ is of Gaussian type with length scale  $L=1$).  }
\label{fig:homo_gau_1}
\end{figure}

Fig. \ref{fig:homo_exp_3} -\ref{fig:homo_gau_1} and Table \ref{table:1} represent the results of twin experiments with different mis-specified (in terms of both amplitude and error correlation) background matrices $\textbf{B}_{\textrm{A}}$, where \cute and \pub are (arbitrarily) applied for 10 iterations.
In each  experiment, the true state $\textbf{x}_t$ is  set to be the shallow water solution at $t=t_f$ in an approximation sub-space defined by the finite difference method. We remind that under the assumption of state-independent prior errors, both the output error and its spatial correlation is independent from the choice of the true state. For the Monte-Carlo validations, 10000 background states are simulated independently following a multivariate Gaussian distribution centred at the true state $\textbf{x}_t$ of fixed background error amplitude with ($\sigma_b=10 \times \sigma_o=0.01 m/s$) and imposed correlation kernel (exponential, Balgovind or Gaussian).  We show explicitly the evolution of assimilation error as well as posterior error correlation (both the exact correlation kernel and the estimation given by \cute and PUB).

More specifically, the  distribution of background error correlations is shown in sub-figures (a) where the exact original error correlation of $\textbf{B}$ (black solid line with triangles) and its estimator ($\textbf{B}_{\textsc{A},n=0}$, green solid line with circles), both being homogeneous and isotropic, are drawn against spatial distance $r$ ($mm$). In order to avoid sampling error for large distance, the error correlation is only considered for $r \in (0,10)$ in a $10 \times 10$ grid. The evolution of average  background/analysis error $||\textbf{x}_t-\textbf{x}_{b,n}||$ in \cute and \pub is shown in sub-figures (b), compared with the analysis error level obtained by a one-shot \textit{3D-VAR} algorithm (the stared green line) and the results of the same \textit{3D-VAR} with the {\em exact} background error covariance matrix (i.e. $\textbf{B}_\textsc{A}=\textbf{B}_\textsc{E}$, represented by the dashed black line). 
The results obtained using the exact background matrix are considered as the optimal target in our study.  We observe in  (b) (Fig. \ref{fig:homo_exp_3}-\ref{fig:homo_gau_1}) that for both proposed approaches, the average values of the analysed error decrease significantly with algorithm iterations. In fact, the first step of \cute and \pub is equivalent to a \textit{3D-VAR} with mis-specified $\textbf{B}_\textsc{A}$ (stared green line). Then, the experiments show that both assimilation errors of \cute (blue curve) and \pub (red curve) decrease and remain stable while approaching better the optimal result (dashed black curve) after a sufficient number of iterations.  

Standard deviations of the estimators are also displayed with transparent shades. Fig. \ref{fig:homo_exp_3} (c) shows
the decrease of the innovation quantity $||y-\textbf{H} \textbf{x}_{b,n}||$. 
 We consider the innovation quantity, available outside the framework of twin experiments, as an appropriate stopping criteria for \cute and \pub algorithms, because of its coherence with the assimilation error (Fig. \ref{fig:homo_exp_3} (a)) both in terms of monotonicity and stability.\\

Despite the fact that output error correlation recognition is significantly improved by \cute and \pub (as shown in Fig. \ref{fig:homo_exp_3} (d-f) and the correlation mismatches in Table (\ref{table:1}), little impact was found on reduction of the output error deviation as shown in the transparent shades in sub-figures (b). The posterior correlation kernels (both the exact one and its estimators), shown in Fig. \ref{fig:homo_exp_3} (d-e) and used to calculate the correlation mismatch in Table \ref{table:1}, are estimated from the data sample by calculating the average correlation value for all pairs of points sharing the same spatial distance in the 2D velocity field of $u$. Correlation kernels obtained in the velocity field of $v$ are very similar. Compared to the prior scenario, with all three initial guess of prior correlation kernel, the bias of the correlation error estimation is significantly reduced \textit{a posteriori}. This improvement is also very noticeable when examining the $L^2$ norm of the correlation mismatches as displayed in Table \ref{table:1}. Sub-figure (f) demonstrates that the AIRM criteria decreases significantly for both approaches after several iterations. It is particularly stable for the \pub method but exhibits some asymptotic non-motonicity for the \cute method.

\noindent

\begin{table}[H]
  \centering
  \resizebox{0.95\linewidth}{!}{%
  \begin{tabular}{ | l|ccc | ccc |}
    \hline
     \multirow{2}{*}{$\textbf{B}_{\textsc{A},n=0}$ kernel choice} &  &   Correlation mismatch ($u$)  &   &   &  {AIRM} & \\
    \cline{2-7}  
     & Initial  & \cute  &  \pub  & Initial  & \cute & \pub\\
    \hline  
    \hline
     {Exponential ($L=3$)} &  0.667 & 0.115 & 0.251 &  28.772
 & 17.510 & 19.069
 \\
     {Balgovind ($L=1$)} &   1.310 & 0.140 &  0.174 &  23.095
 & 15.607 & 15.116 \\
     {Gaussian ($L=1$)} &   1.834 & 0.305 &  0.660 &  26.642
 & 19.518 & 20.957
 \\
      \hline
  \end{tabular}}
  \caption{Quantification of results of the \cute and \pub iterative methods in terms of error correlation identification at the tenth iteration. The prior error covariance $\textbf{B}_\textsc{E}$ is set to be of  Balgovind type with correlation length $L=2$, homogeneous and state-independent. The mismatch of calibrated correlation functions is calculated with an $L^2$ norm error between the one-dimensional correlation curves. The AIRM criteria is also reported.  
  }
  \label{table:1}
\end{table}

In conclusion, the iterative approaches improve the assimilation both in terms of  reliability of the analyzed error covariance estimate as well as accuracy of the analyzed state. Nevertheless, the final result seems to remain dependent, to some extent, to the level of dissimilarity between the initial guess $\textbf{B}_\textsc{A}$ and the exact $\textbf{B}$. In particular, poorer results are obtained when the prior background error correlation distributions are extremely misspecified (e.g. Gaussian($L=1$)), regardless of the type of correlation kernel structures considered.

All these numerical results and analyses are obtained under the assumption of a high level of background error variance amplitude, which are here under-estimated by the assumed covariance matrix $\textbf{B}_\textsc{A}$ (i.e. $Tr (\textbf{B}_\textsc{A})< Tr(\textbf{B}_\textsc{E})$). This assumption is consistent with the phenomenon of background error inflation as mentioned in \ref{sec:Introduction to 3D-Var}.  \rot{We remind that as the dimension of the observation space is inferior to the one of the state space, the equation
\begin{align}
    \textbf{y} = \textbf{H} \textbf{x} 
\end{align}
is underdetermined. It thus defines a hyperplane in the space of $\textbf{x}$. Fig. \ref{fig:relative_exp_3} (b) shows that the \cute method converges to a stable state when the assimilation error $||\textbf{x}_{a,n}^\textrm{\cute} - \textbf{x}_t||$ is very close to the optimal target $||\textbf{x}_{a}^\textrm{optimal} - \textbf{x}_t||$. However, we don't necessarily have
\begin{align}
    \textbf{x}_{a,n} \rightarrow \textbf{x}_{a}^\textrm{optimal}.
\end{align}
}

\subsection{Twin experiments with state-dependent prior errors } \label{subsec:relative_error}
In this section, we are interested in  the performance of our  methods in the case of state-dependent errors, i.e. when the assumption of independence between the true state and estimation errors no longer stands (such as the optimal property of the maximum likelihood).
State-dependent uncertainties are certainly more complex but it is more realistic for numerous industrial applications. Very recent effort was given along this path in order to improve  data assimilation algorithms, e.g. (\cite{Bishop2018}).  \\

As for the case of homogeneous prior errors, background states and observations are simulated by Gaussian distributions centred around true values (i.e. respectively $\textbf{x}_t$ and $H \textbf{x}_t$ for background states and observations). However, the standard deviation at each coordinate is set to be proportional to the magnitude of the true state, while keeping the prior correlation structures as described in \ref{eq:Gaussian}.\\
In order to better define how state-dependent prior errors are simulated, we denote $\textbf{D}^B$ (resp. $\textbf{D}^R$) as the diagonal of the exact covariance matrix $\textbf{B}_\textsc{E}$ (resp. $\textbf{R}_\textsc{E}$). Above assumption of state-dependent errors leads to: 
\begin{align}
    \textbf{D}_i^B& = (\mu_b \times \textbf{x}_{t,i})^2 \label{eq:diag_B} \\ 
    \textbf{D}_i^R& = (\mu_o \times (\textbf{H}\textbf{x}_t)_i)^2,  \label{eq:diag_R}%\notag \label{eq:diag_R}
\end{align}
where $i$ refers to an index mapping to the two dimensional fields and $(\mu_b, \mu_o)$ stand for two real coefficients. Combining this state-dependent variance and a homogeneous structure of error correlation, state-dependent error covariance matrices can be written as:
\begin{align}
    \textbf{B}_\textsc{E}&=(\textbf{D}^B)^{\frac{1}{2}} \textbf{Cor}_{B,\textsc{E}} (\textbf{D}^B)^{\frac{1}{2}}  \label{eq:correlation_covariance}\\
    \textbf{R}_\textsc{E}&=(\textbf{D}^R)^{\frac{1}{2}} \textbf{Cor}_{R,\textsc{E}} (\textbf{D}^R)^{\frac{1}{2}}, \notag 
\end{align}
where the exact prior correlation matrices $\textbf{Cor}_{B,\textsc{E}}$ and $\textbf{Cor}_{R,\textsc{E}}$ are still chosen to follow homogeneous and isotropic correlation kernels as in the case of  \ref{subsec:homo_error}. 

The two velocity fields $u$ and $v$ are supposed to be uncorrelated in terms of prior estimation error. Thus both $\textbf{B}_\textsc{E}$ and $\textbf{B}_\textsc{A}$ follow a block diagonal structure. This is obviously a very  crude assumption in the context of incompressible fluid mechanics systems. Since observation errors are also state-dependent in this case, the associated observation error covariance cannot be known exactly \textit{a priori}. We introduce the notation $R_\textsc{A}$ for assuming observation error covariance, which is different from the true observation error covariance only in this section (\ref{subsec:relative_error}). The assumption of relatively higher background error is also respected by setting $10 \%$ standard deviation for background state (i.e. $\mu_b=10 \%$) while $1 \%$ for observations (i.e. $\mu_o=1 \%$) in twin experiments.

\begin{figure}[!ht]
    \begin{subfigure}[H]{.3\textwidth}
      \includegraphics[trim=2cm 8cm 2cm 8.8cm,clip=true,width=2.2in]{Fig9.pdf}
      \subcaption{}
    \end{subfigure}
  \centering
    \begin{minipage}[c]{.3\textwidth}
      \includegraphics[width=2.in]{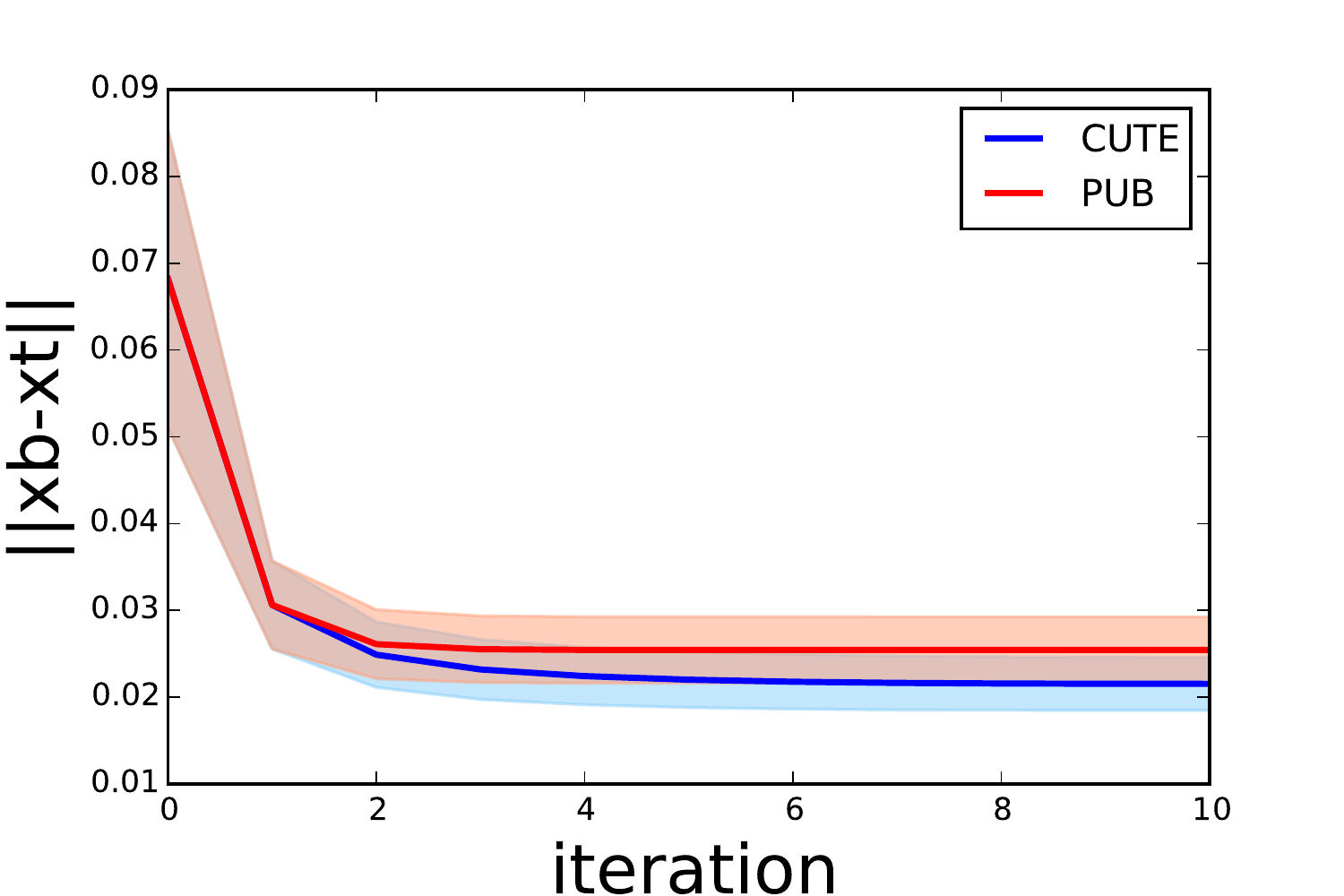}
      \subcaption{}
   \end{minipage}
    \begin{subfigure}[H]{.3\textwidth}
      \includegraphics[trim=2cm 8cm 2cm 8.8cm,clip=true,width=2.2in]{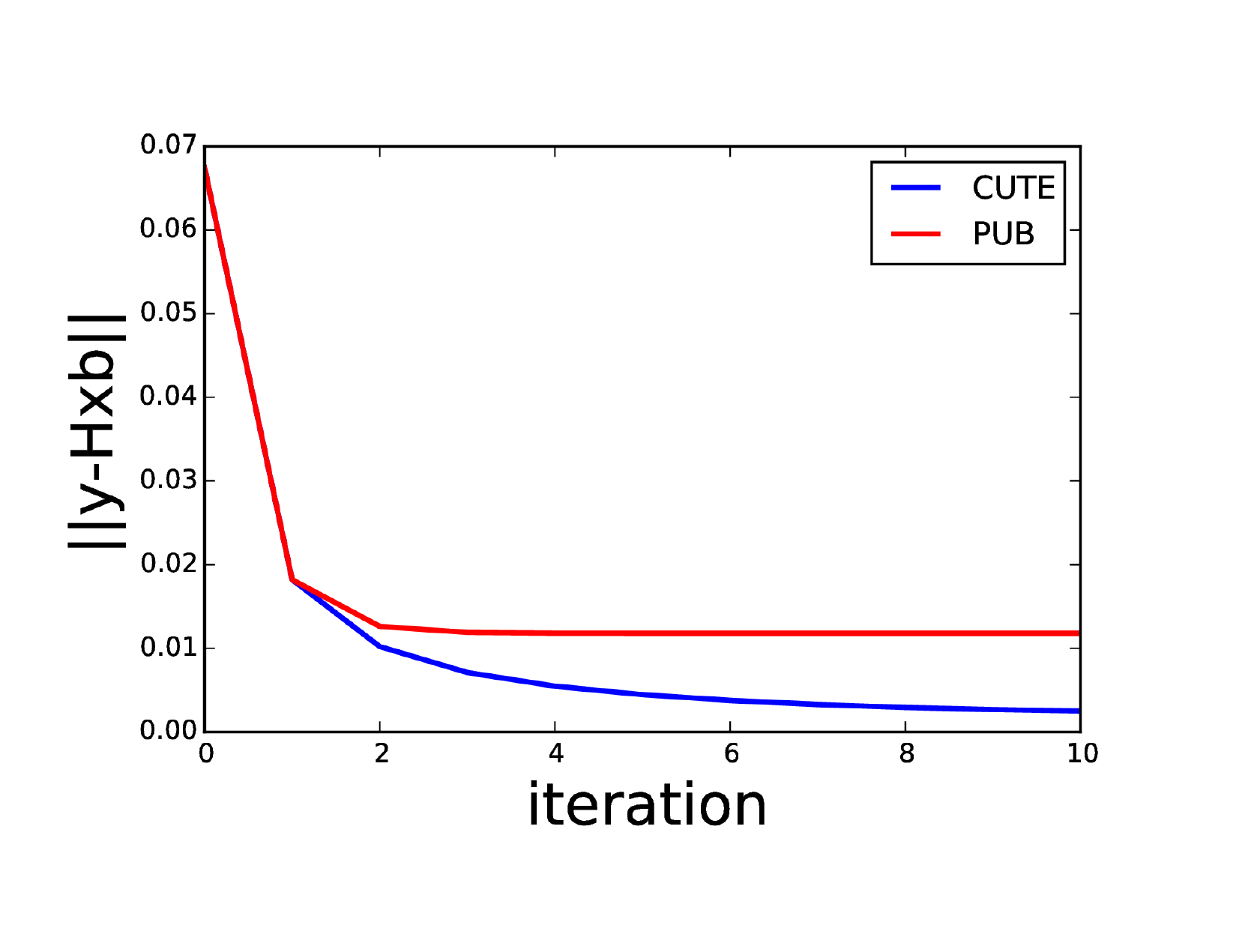}
      \subcaption{}
    \end{subfigure}        
    \begin{subfigure}[H]{.3\textwidth}
      \includegraphics[trim=2cm 8cm 2cm 8.8cm,clip=true,width=2.2in]{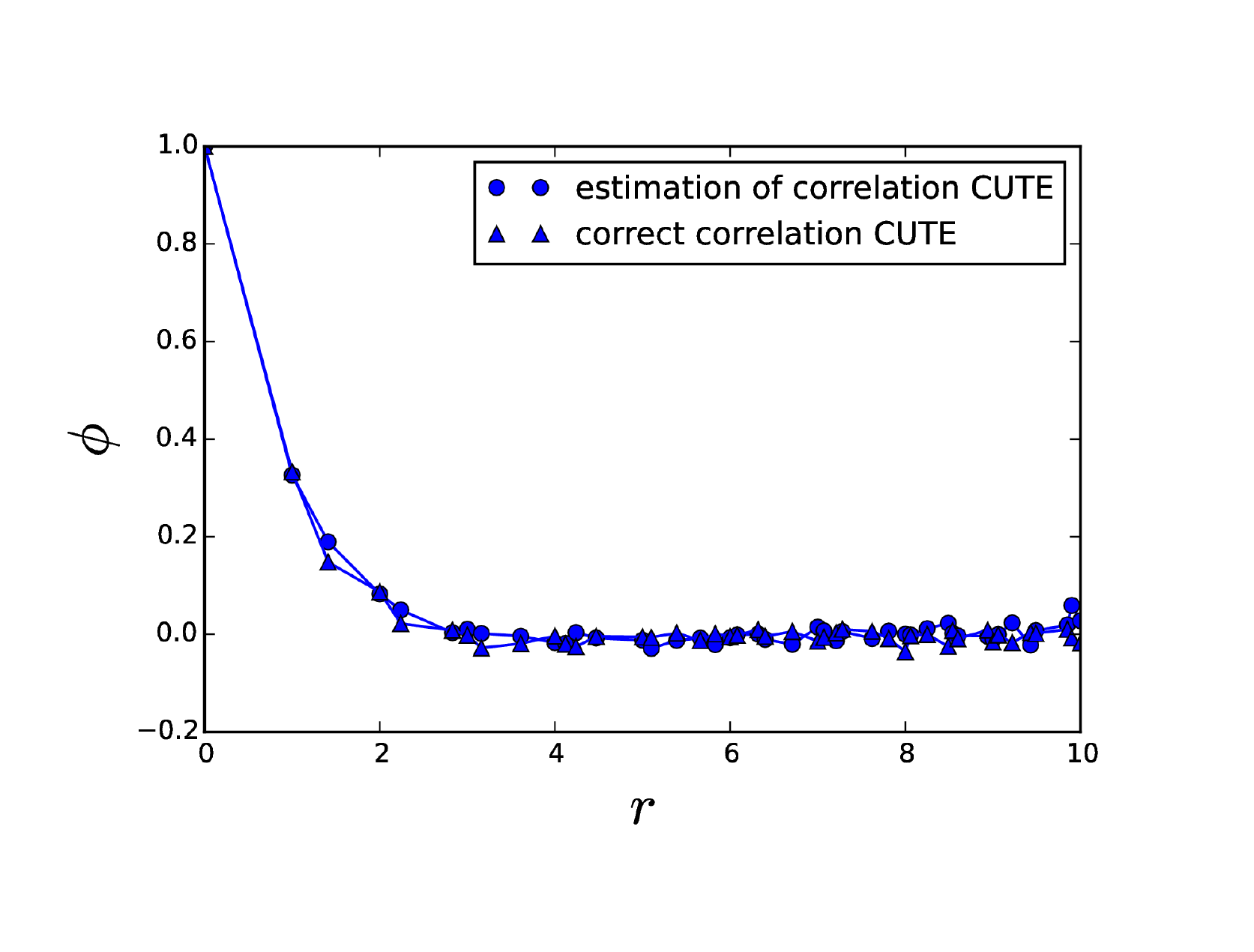}
      \subcaption{}
    \end{subfigure}
    \begin{subfigure}[H]{.3\textwidth}
      \includegraphics[trim=2cm 8cm 2cm 8.8cm,clip=true,width=2.2in]{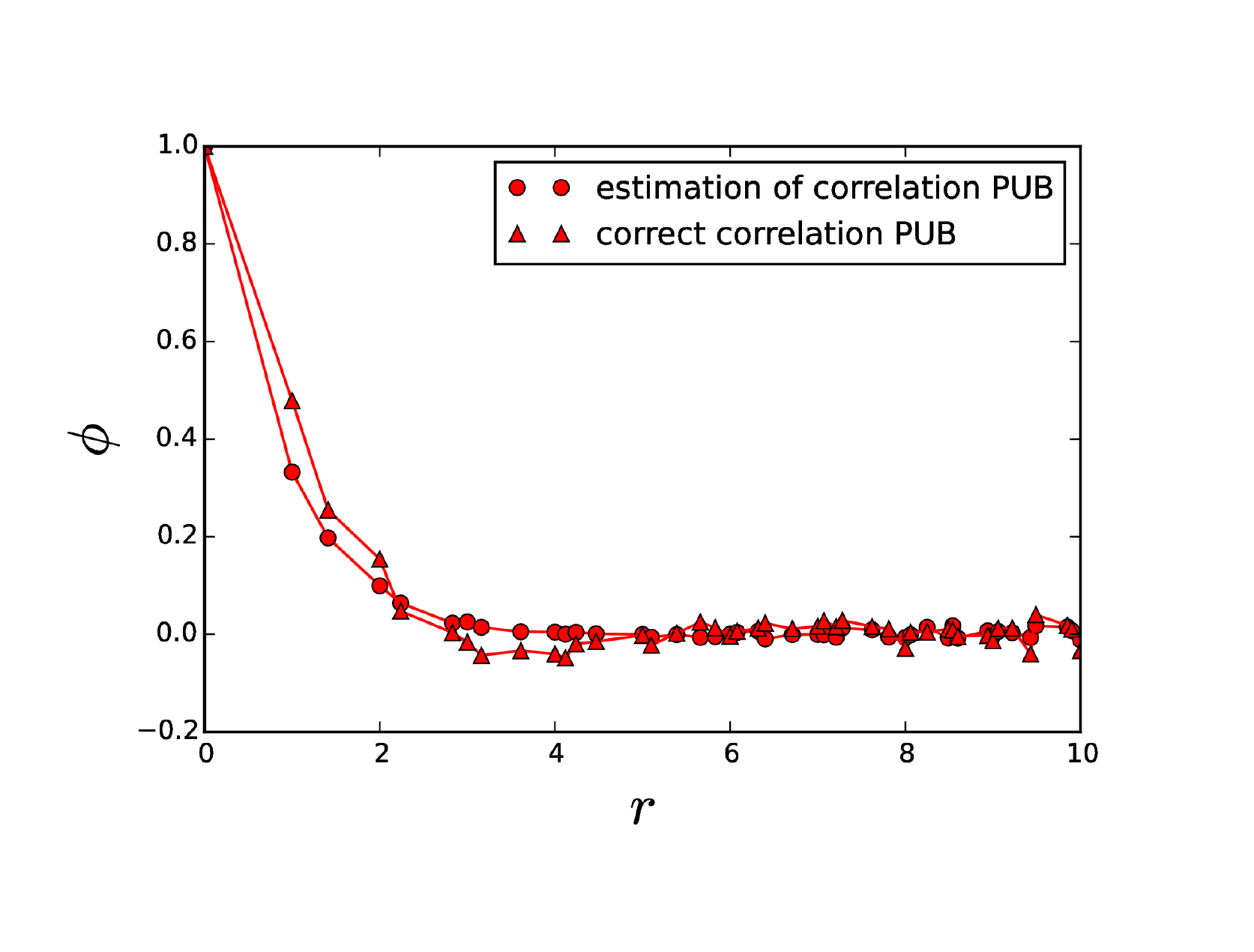}
      \subcaption{}
    \end{subfigure}
    \begin{subfigure}[H]{.3\textwidth}
      \includegraphics[trim=2cm 8cm 2cm 8.8cm,clip=true,width=2.2in]{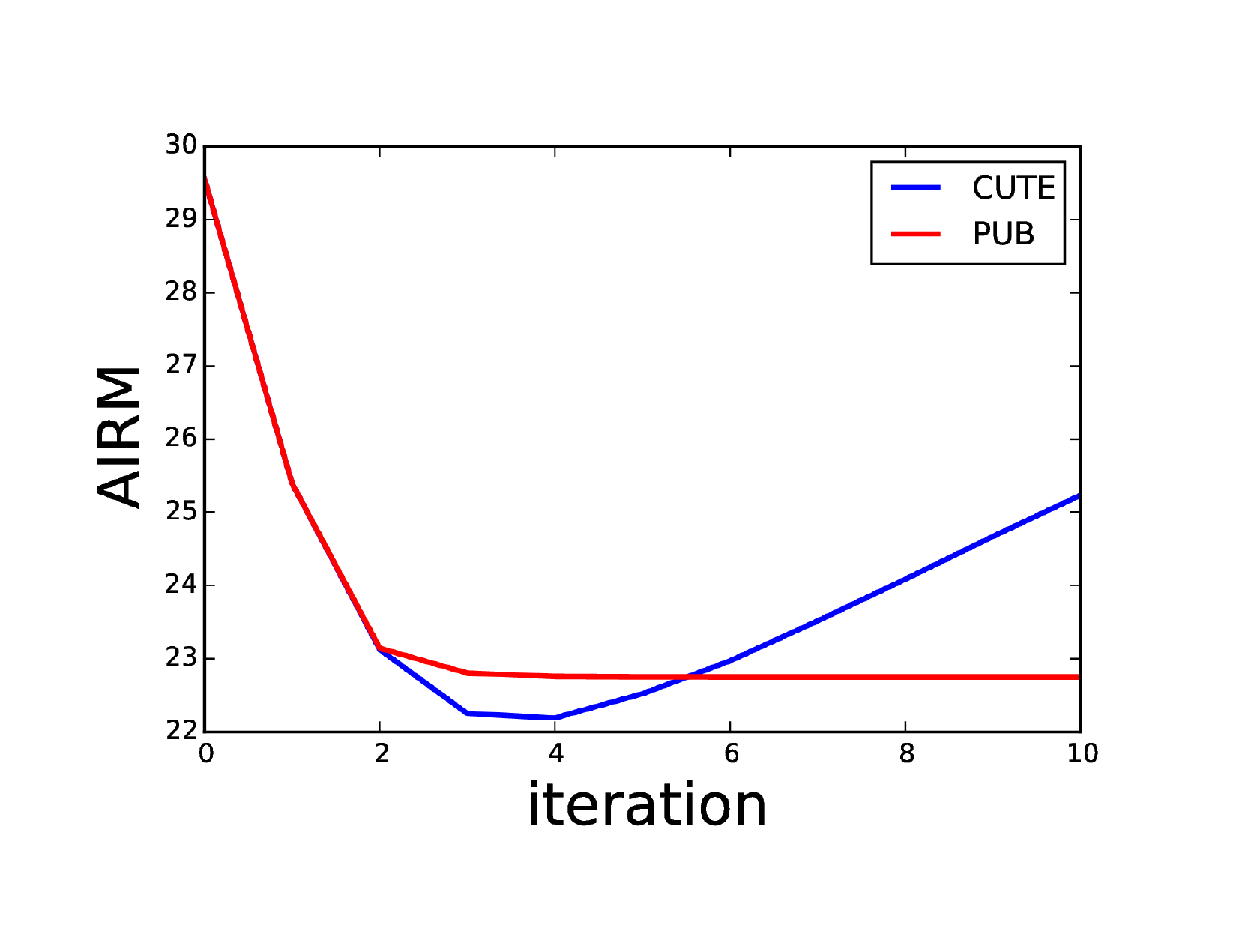}
      \subcaption{}
    \end{subfigure}  
\caption{Twin experiments with state-dependent prior error for background state and observations. Figures on the first line refer to the initial choice of prior correlations (a) and the evolution of assimilation error (b) and innovation quantities (c), while figures on the second line monitor iterated quantities extracted from the errors covariance in the velocity field of $u$(d-f).  In this test, the correlation of $\textbf{B}_{\textsc{A},n=0}$ is chosen to follow an exponential kernel with $L=3$ (shown by the green curve in (a)).}

\label{fig:relative_exp_3}
\end{figure}

\begin{figure}[!ht]
  \centering
    \begin{subfigure}[H]{.3\textwidth}
      \includegraphics[trim=2cm 8cm 2cm 8.8cm,clip=true,width=2.2in]{Fig15.pdf}
      \subcaption{}
    \end{subfigure}
    \begin{minipage}[c]{.3\textwidth}
      \includegraphics[width=2.in]{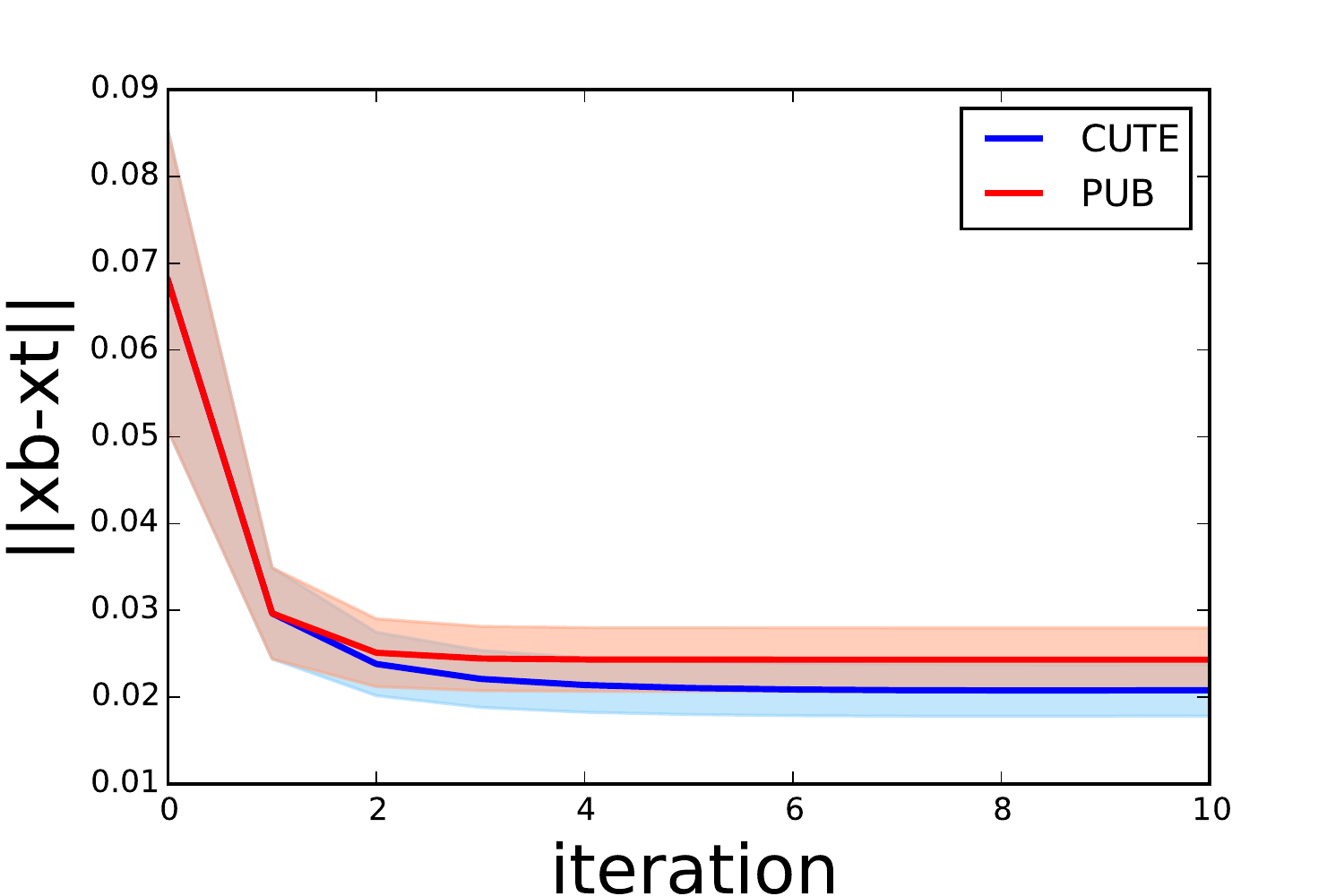}
      \subcaption{}
   \end{minipage}
\caption{Evolution of assimilation error in twin experiments using same simulated observations as Fig. \ref{fig:relative_exp_3}  with different initial background matrix estimation (the correlation kernel of $\textbf{B}_{\textsc{A},n=0}$ is of Balgovind type with length scale  $L=1$).  }
\label{fig:relative_bal_1}
\end{figure}

\begin{figure}[!ht]
  \centering
      \begin{subfigure}[H]{.3\textwidth}
      \includegraphics[trim=2cm 8cm 2cm 8.8cm,clip=true,width=2.2in]{Fig17.pdf}
      \subcaption{}
    \end{subfigure}
    \begin{subfigure}[H]{.3\textwidth}
      \includegraphics[width=2.in]{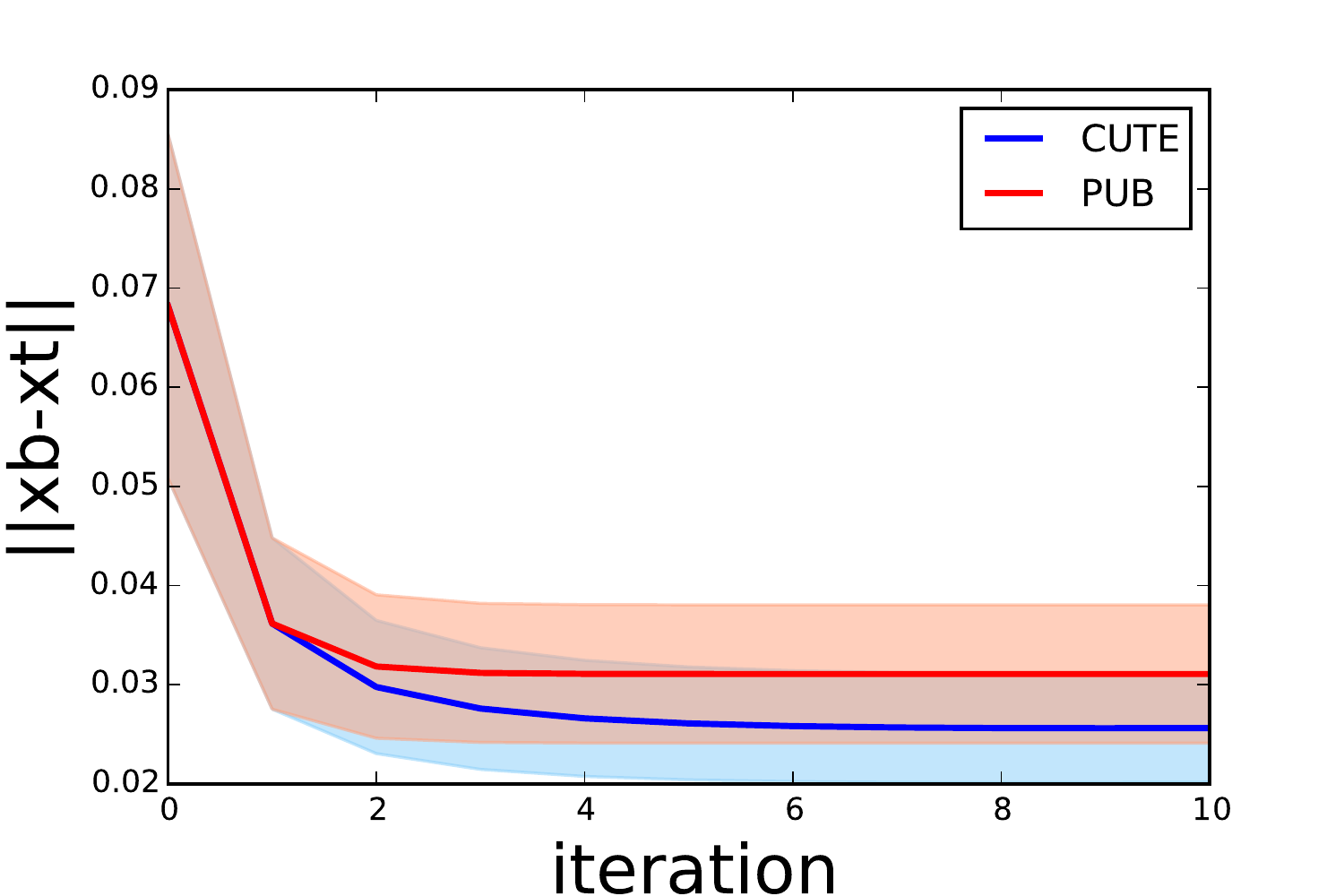}
      \subcaption{}
   \end{subfigure}
\caption{Evolution of assimilation error in twin experiments using same simulated observations as Fig. \ref{fig:relative_exp_3}  with different initial background matrix estimation (the correlation kernel of $\textbf{B}_{\textsc{A},n=0}$ is of Gaussian type with length scale  $L=1$).  }

\label{fig:relative_gau_1}
\end{figure}

Monte Carlo twin experiments of 10000 tests with state-dependent prior errors are carried out as presented in Fig. \ref{fig:relative_exp_3}-\ref{fig:relative_gau_1}. We keep homogeneous structure of the assumed covariance matrices $\textbf{B}_\textsc{A}$ (constructed using correlation kernels) and $R_\textsc{A}$ (set to be the identity matrix) as in section \ref{subsec:homo_error}. We also choose to keep the trace of $\textbf{B}_\textsc{A}$ and $\textbf{R}_\textsc{A}$ during iterative processes CUTE/PUB.
Results in Fig. \ref{fig:relative_exp_3}-\ref{fig:relative_gau_1} and Table \ref{table:2} show that for state-dependent errors, \cute and \pub iterative methods could also significantly reduce  the output errors (sub-figures (b) of Fig. \ref{fig:relative_exp_3}-\ref{fig:relative_gau_1}) compared to the first iteration (standard \textit{3D-VAR} algorithm), as well as the innovation quantity (sub-figures (c)). The latter remains an appropriate candidate for the stopping criteria. Important improvements are obtained in terms of decreasing the bias of error correlation estimation as shown in sub-figures $(e)$ and $(f)$ (comparing with $(a)$). However, as shown in Fig. \ref{fig:relative_exp_3} (f) and in the last two columns in Table \ref{table:2}, the AIRM criteria which monitors a global correlation matrix estimation mismatch, reveals a risk of saturation of the use of the same observation data set for the \cute method after a certain number of iterations. This is due to the imperfect knowledge of observation error covariance. The \pub method is less sensitive and more stable in this case. However, as for the case of state-independent prior errors, \cute owns a slight advantage over \pub in terms of assimilation error reduction.

\begin{table}[H]
  \centering
  \resizebox{0.95\linewidth}{!}{%
  % \begin{tabular}{|P{2.5cm}|P{2.5cm}|P{2.5cm}|P{2.5cm}|}
  \begin{tabular}{ | l|ccc | ccc |}
    \hline
     \multirow{2}{*}{$\textbf{B}_{\textsc{A},n=0}$ kernel choice} &  &   Correlation mismatch($u$) &   &   &  {AIRM} & \\
    \cline{2-7}  
     & Initial  & \cute  &  \pub  & Initial  & \cute & \pub\\
    \hline  
    \hline
     {Exponential ($L=3$)} &  0.586 & 0.147 & 0.220 &  29.550
 & 25.234 & 22.752
 \\
     {Balgovind ($L=1$)} &   1.191 & 0.207 &  0.180 &  24.181
 & 24.439 & 19.785 \\
     {Gaussian ($L=1$)} &   1.733 & 0.333 & 0.662 & 27.495
 & 28.068 & 23.569
 \\
      \hline
  \end{tabular}}
   \caption{Quantification of assimilation results of the \cute and \pub iterative methods in terms of error correlation identification at the tenth iteration. The prior error covariance is set to be non homogeneous and state-dependent with correlation matrix of Balgovind type, $L=2$. The mismatch of calibrated correlation functions is calculated with an $L^2$ norm error between the calibrated one-dimensional correlation curves. The AIRM criteria is also reported. 
  }
  \label{table:2}
\end{table}

\subsection{Twin experiments in a successive data assimilation process of reconstruction/prediction}\label{sec:dym_twin}

\begin{figure}[!ht]
\centering
    \begin{subfigure}[H]{.4\textwidth}
      \includegraphics[trim=0cm 6.5cm 0cm 8cm,clip=true,width=2.5in]{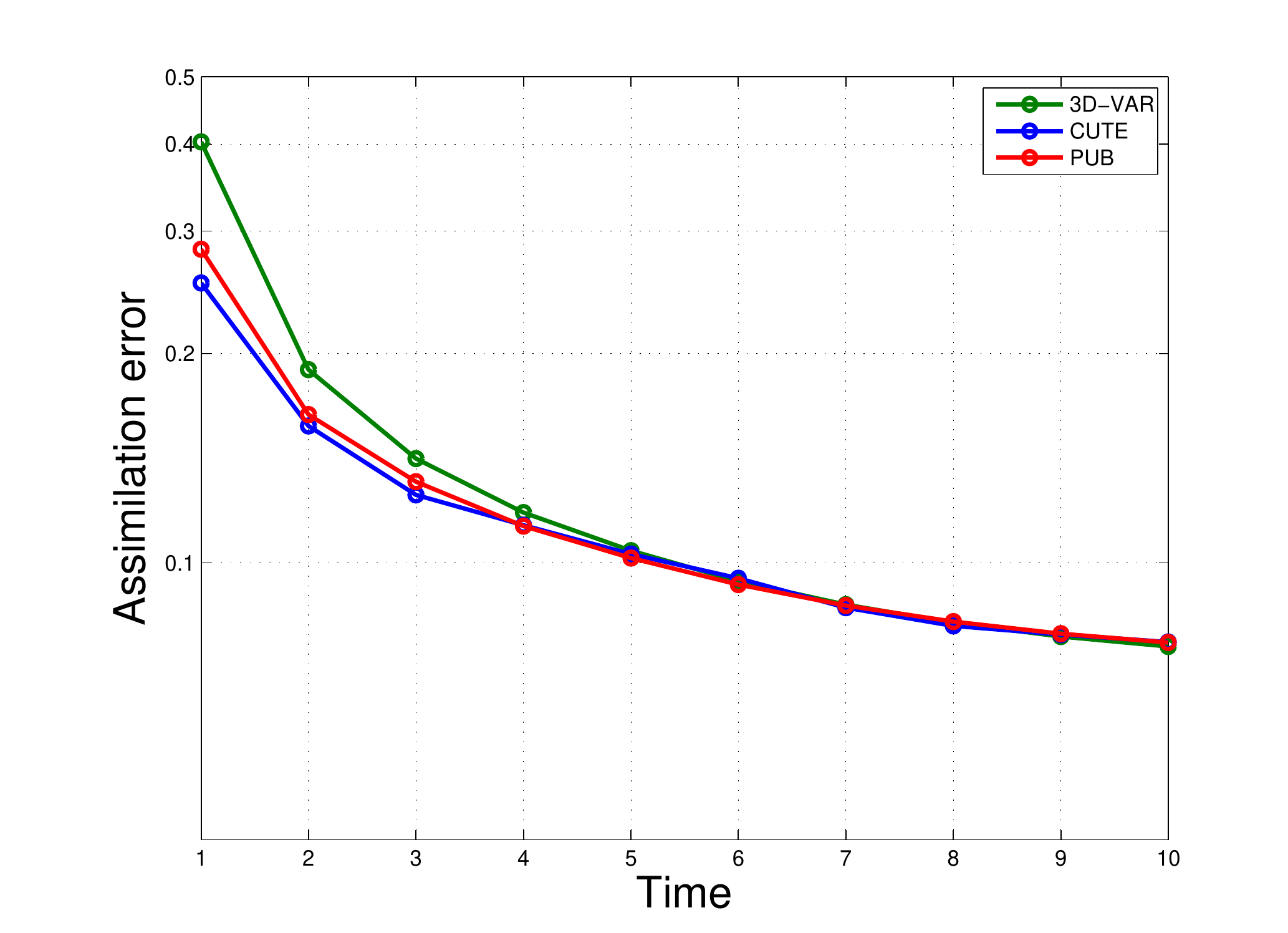}
      \subcaption{$\sigma_b=10 \sigma_o$}
    \end{subfigure}
    \begin{subfigure}[H]{.4\textwidth}
      \includegraphics[trim=0cm 6.5cm 0cm 8cm,clip=true,width=2.5in]{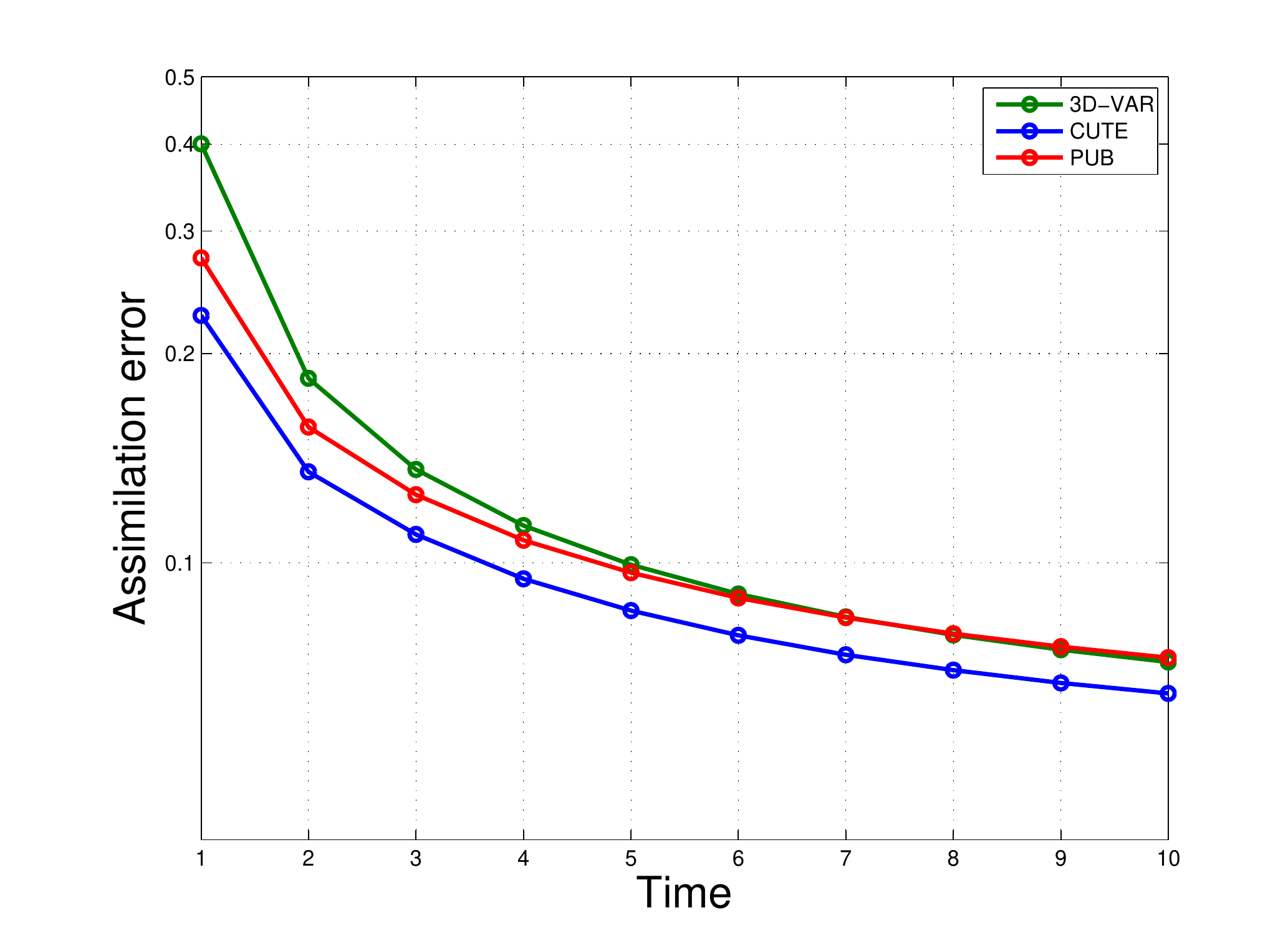}
      \subcaption{$\sigma_b=100 \sigma_o$}
    \end{subfigure}
\caption{Comparison of standard \textit{3D-VAR} method with iterative methods in terms of evolution of assimilation error in a dynamical twin experiments framework, where a data assimilation reconstruction takes place every $2 \times 10^{-3}s$. Semi-log grid is used for ordinate coordinates. In these experiments, iterative methods \cute and \pub are only applied at the first reconstruction of the process, followed later by standard  \textit{3D-VAR}. Simulations are made based on two different level of prior background-observation error: $\sigma_b=10 \sigma_o$ (a), $\sigma_b=100 \sigma_o$ (b). }
\label{fig:only_first_dym}
\end{figure}

The idea of this section is to anticipate on the use of these types of approaches in the wider framework of time-dependent data assimilation problems.
Based on the shallow water propagation model introduced in section \ref{subsec:relative_error}, we construct new twin experiments of a dynamical field reconstruction and prediction relying on successive applications of data assimilation algorithm using flow-independent background matrix $\textbf{B}_\textsc{A}$. The choice of the test model is made for its simplicity and for better revealing the impact of \cute and \pub methods.  The state dimension remains $200$ which is composed of two squarely meshed velocity fields of $10 \times 10$ each as for static reconstruction with state-independent prior errors in \ref{subsec:homo_error}. In order to focus on the impact of background error propagation, correct boundary conditions are simulated independently in an error free framework and provided at each reconstruction step for state-transition model  to avoid an overlay of model resolution error. In order to observe the impact of \cute and \pub methods in a long term data assimilation procedure, we choose to apply solely \cute and \pub at the first reconstruction step of the process for a fixed number of iterations $n$ ($n=10$ in following numerical tests), following \textit{3D-VAR} reconstructions every $2 \times 10^{-3} s$. With a significant improvement of assimilation error reduction and error correlation recognition provided by \cute or PUB, this advantage should be recognised and kept by a standard variational method (in our case, the \textit{3D-VAR} method) for several further steps in a data assimilation chain.  We then compare the results obtained by a standard approach of \textit{3D-VAR} all the way along. We remind that the difference among the three data assimilation processes shown in Fig. \ref{fig:only_first_dym} are only the first reconstruction at $t=10^{-3}s$.  The evolution of average assimilation error of $100$ independent dynamical simulations is illustrated in Fig. \ref{fig:only_first_dym} with two different levels of initial errors. We observe a significant improvement due to the iterative process of \cute and \pub at the first several reconstruction steps. The gaps among the three curves then tend to disappear. In fact, even starting with a high level of noise, a standard successive data assimilation process should be capable of providing a reasonably good long term prediction for both assimilation error reduction and covariance recognition when the information about the state-transition model is accurate enough (\cite{Rabier2005}), which is the case in our experiments. 

\begin{figure}[!ht]
  \centering
    \begin{subfigure}[H]{.4\textwidth}
      \includegraphics[trim=0cm 6.5cm 0cm 8cm,clip=true,width=2.5in]{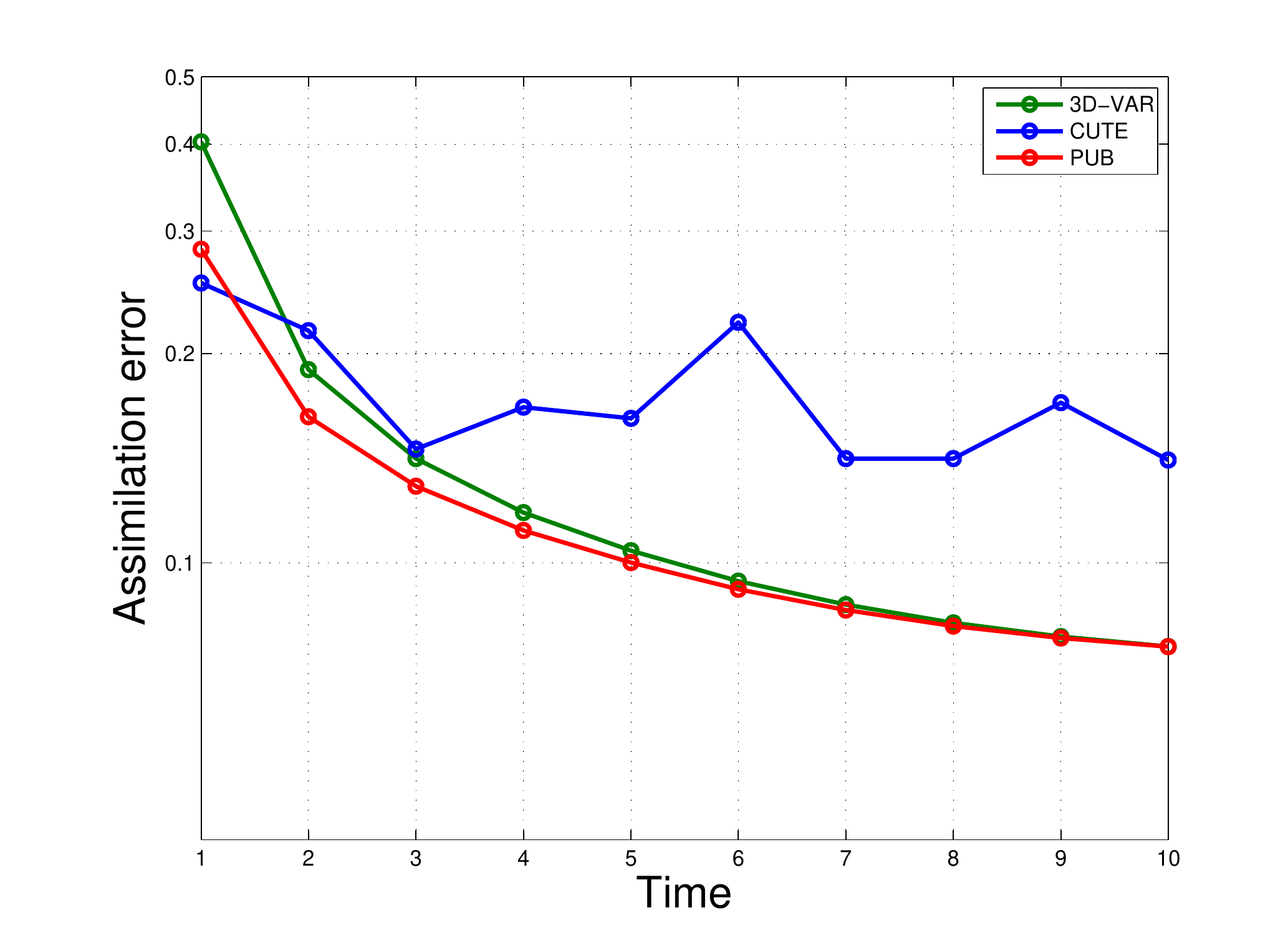}
      \subcaption{$\sigma_b=10 \sigma_o$}
    \end{subfigure}
    \begin{subfigure}[H]{.4\textwidth}
      \includegraphics[trim=0cm 6.5cm 0cm 8cm,clip=true,width=2.5in]{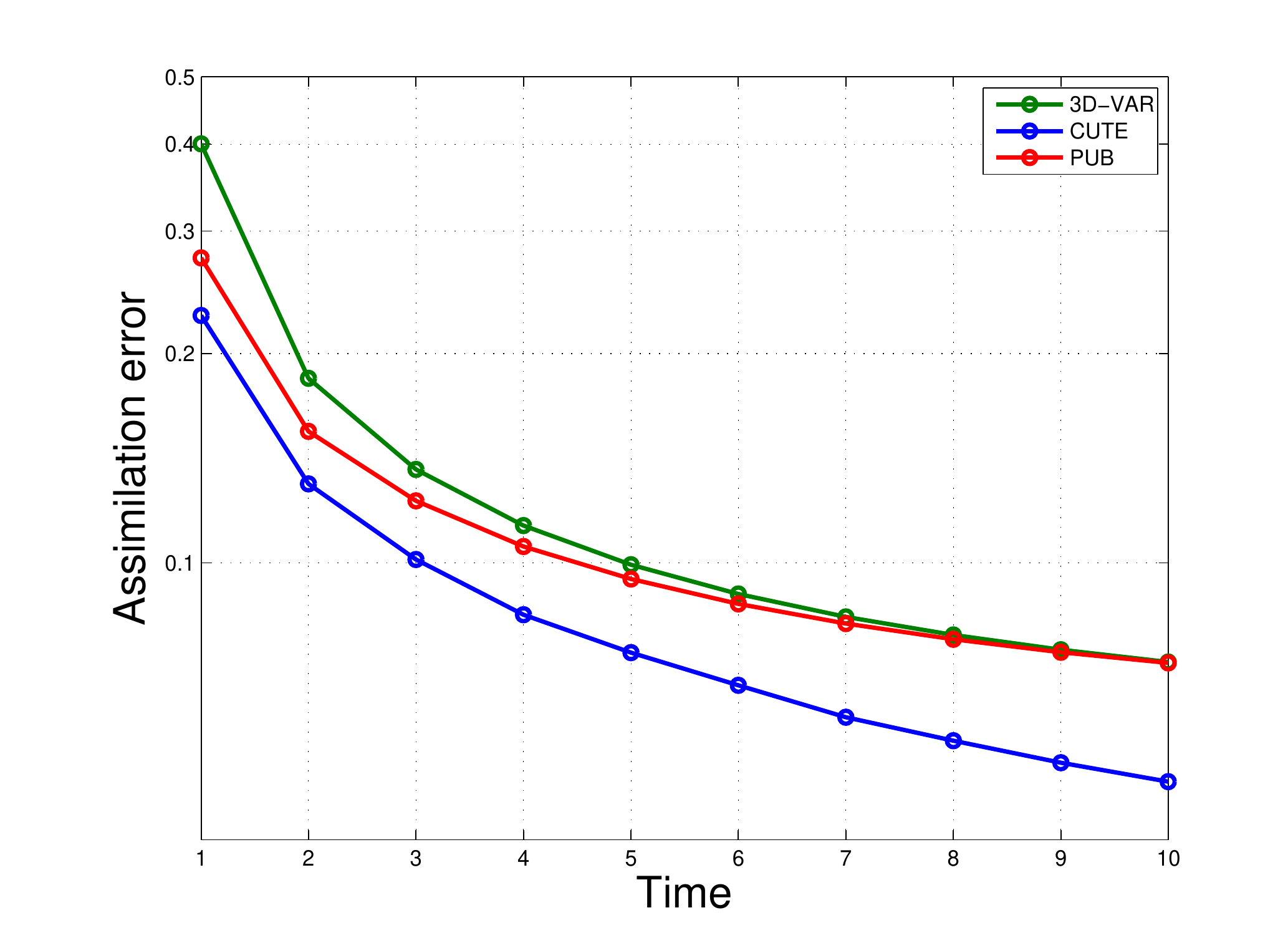}
      \subcaption{$\sigma_b=100 \sigma_o$}
    \end{subfigure}
\caption{Comparison of {\em dynamical} twin experiments, with same initial conditions as in Fig. \ref{fig:only_first_dym}, where ten \cute and \pub sub-iterations are performed at each assimilation step, illustrated by the green and red disk symbols. }
\label{fig:all_dym}
\end{figure}

The result in Fig. \ref{fig:only_first_dym} confirms the interest of applying CUTE, \pub methods for a short term prediction. It is also shown that when the assumption of high level or inflated background errors is well respected (Fig. \ref{fig:only_first_dym} (b)), the "advantage" of an iterative process at the initial step could be kept longer in the dynamical assimilation. However, when the observation error is not sufficiently negligible relatively to the background error, a continuous correction by iterative processes is helpful. The same holds when the information about the dynamical state-transition is not precise, especially in a highly nonlinear system where the misspecification of estimation errors could be enlarged in the successive predictions.
Therefore, in these cases an interest can be arisen to apply the iterative methods continuously at several different moments. We present in Fig. \ref{fig:all_dym}, the same dynamical twin experiments with an implementation of \cute, \pub at each assimilation step (i.e. every $2 \times 10^{-3}s$) instead of \textit{3D-VAR}. By construction, the first steps (both (a) and (b)) of Fig. \ref{fig:only_first_dym} and \ref{fig:all_dym} are equivalent.

We observe clearly from Fig. \ref{fig:all_dym} that, when the observation error is negligible compared to the background error, the implementation of \cute method at each assimilation step enables a continuous reduction of assimilation error. On the other hand, consistent with previous analysis, the \cute method is very sensitive to the level of observation errors, especially when being reapplied several times in a dynamical procedure with a good knowledge of the transition model (no inflation of background errors) as presented in Fig. \ref{fig:all_dym} (a). In general, the  \pub method remains more robust and less sensitive to the hypothesis of the high level of background error. 

\section{Conclusion}

In this paper we introduced two novel data assimilation iterative methods recycling the observation data for the purpose of damping the detrimental effect of a poor knowledge of the background error covariance. In this framework, we have shown that a naive approach which neglects the background-observation correlation introduced by the iterative process is prone to failure. This indicates that there is a need for a complete covariance updating, as being carried out in the proposed approaches. 

Under the assumption of perfect knowledge of the observation error covariance and the transformation operator,  we numerically demonstrated  that CUTE and PUB methods could noticeably improve output error correlation identification as well as reduce the assimilation error for a variety of initial guesses of the background error covariance matrix,  when prior errors are either state-independent or state-dependent. These two methods are different from other iterative methods, in the sense that they not only update the variance of state components but also the background state correlation structure. \rtwo{Other covariance tuning methods, such as the full Desroziers diagnostic used  in the observation space, require more data especially for large-scale problem.}\\ Originally developed for the purpose of statistical reconstructions or short term predictions,
we have shown that there might be an interest in reapplying the proposed algorithms several times in a dynamical assimilation chain. Limitations of these two methods have also been pointed out in this article, in particular, concerning the risk of straining a redundant observation data set without a careful monitoring of the convergence results.

\rot{The difference between \cute and \pub resides mainly in the minimization function, where the covariance between updated background and observation 
is only taken into account into the \pub method. This feature makes the \pub method more robust, i.e. less sensitive to the assumption of the trace of the prior errors $Tr(\textbf{B}_\textrm{E}) > Tr(\textbf{R}_\textrm{E})$ and the usage of the same observation data set.  
Numerically, we have also found that the performance of \cute can be more optimal when the background error is much underestimated. In fact, the estimation of $COV(\epsilon_b, \epsilon_y)$ is also based on the prior knowledge of matrix $\textbf{B}$. In summary, we recommend  the utilisation of CUTE when the initial background error covariance matrix is set arbitrarily, especially when it is probably underestimated while the PUB method can be more appropriate when limited data are available for making a rough estimation of $\textbf{B}$ or its diagonal.}

In terms of computational cost, \cute and \pub methods can be relatively more expensive than the Desroziers approach which only requires \textit{a posteriori} computation of matrix traces. However, when no linearization of the observation operator is needed, the updating process can be  done  aside once the initial guess for covariance matrices are available and independently from the current background state. This feature  promotes a more flexible use of these methods with much lower computational overheads. Future work will investigate along this path of research and will access  their performance in a more realistic/sophisticated industrial application case.

\bibliographystyle{abbrv}      
% basic style, author-year citations
\bibliography{serra}
\end{document}